\documentclass[twocolumn,superscriptaddress,showpacs]{revtex4}
\usepackage[british,UKenglish,USenglish,english,american]{babel}
\usepackage{amsmath}
\usepackage{amssymb}
\usepackage{latexsym}
\usepackage{enumerate}
\usepackage{amssymb}
\usepackage{graphicx}
\usepackage{slashed} 

\begin{document}
\title{Timelike Compton scattering off the neutron }

\author{M. Bo\"er$^1$, M. Guidal$^1$ and M. Vanderhaeghen$^2$ \\
\textit{$^1$Institut de Physique Nucl\'eaire d'Orsay, CNRS-IN2P3, Universit\'e Paris-Sud, 
Universit\'e Paris-Saclay, 91406 Orsay, France}\\
\textit{$^2$Institut f\"ur Kernphysik and PRISMA Cluster of Excellence,  Johannes Gutenberg Universit\"at, Mainz, Germany.}
}

\begin{abstract}
We study the exclusive photoproduction of an electron-positron pair
on a neutron target in the Jefferson Lab energy domain. The reaction
consists of two processes: the Bethe-Heitler and the Timelike Compton Scattering.
The latter process provides potentially access to the 
Generalized Parton Distributions (GPDs) of the nucleon. We calculate all
the unpolarized, single- and double-spin observables of the reaction
and study their sensitivities to GPDs.
\end{abstract}

\maketitle

\section{Introduction}

In a recent article~\cite{Boer:2015hma}, we studied the exclusive photoproduction 
of an electron-positron pair off a proton target, i.e. the $\gamma p\to p' e^+e^-$ reaction, in the multi-GeV beam energy domain. For sufficiently large invariant masses of the final lepton pair 
$Q'^2=(e^++e^-)^2\gtrsim$ 4 GeV$^2$ and small squared nucleon momentum transfer $-t=-(p'-p)^2\lesssim$ 1 GeV$^2$, the process
allows to study the partonic substructure of the proton. In particular, it provides
access to the Generalized Parton Distributions (GPD) of the proton. We refer 
the reader to Refs.~\cite{Goeke:2001tz,Diehl:2003ny,
Belitsky:2005qn,Guidal:2013rya,Boffi:2007yc} for reviews and details on 
the GPD concepts and formalism. In simple terms, the
GPDs are universal structure functions which allow to map the correlated
transverse position-longitudinal momentum distributions of the quarks and gluons 
within the nucleon. These correlations are up to now barely known. 
In Ref.~\cite{Boer:2015hma}, we calculated, in addition to the unpolarized cross sections, 
all the single-and double-spin beam/target observables of the 
$\gamma p\to p' e^+e^-$. We showed their sensitivities to the different GPDs.
In this article, we extend this work to a neutron target with the aim of studying
the sensitivity of the process to neutron GPDs. 

\begin{figure}[htbp]
\begin{center} 
\includegraphics[width=8.5cm,height=5.2cm]
{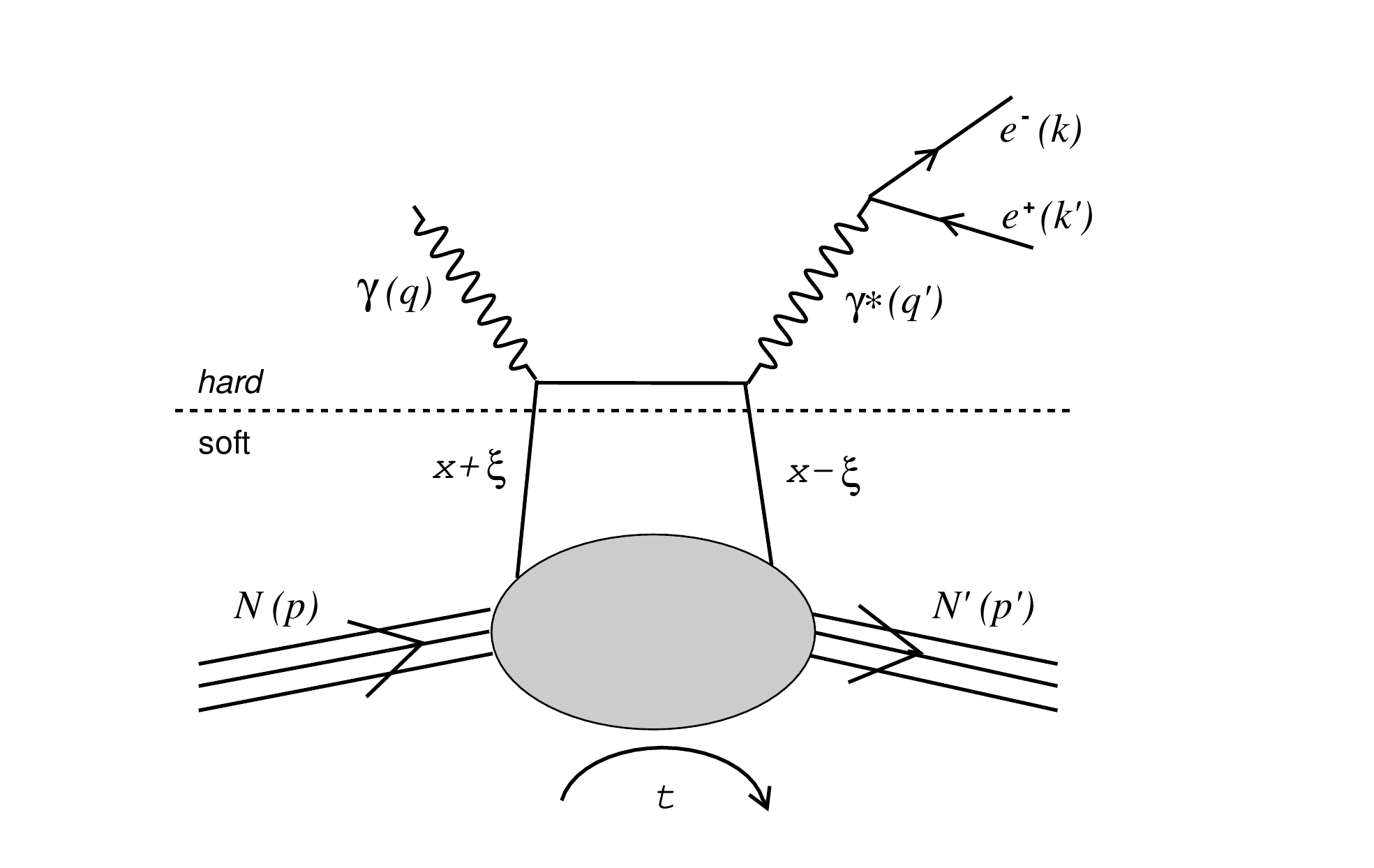} \\ \vspace*{-0.8cm}
\includegraphics[width=7cm,height=5.2cm]
{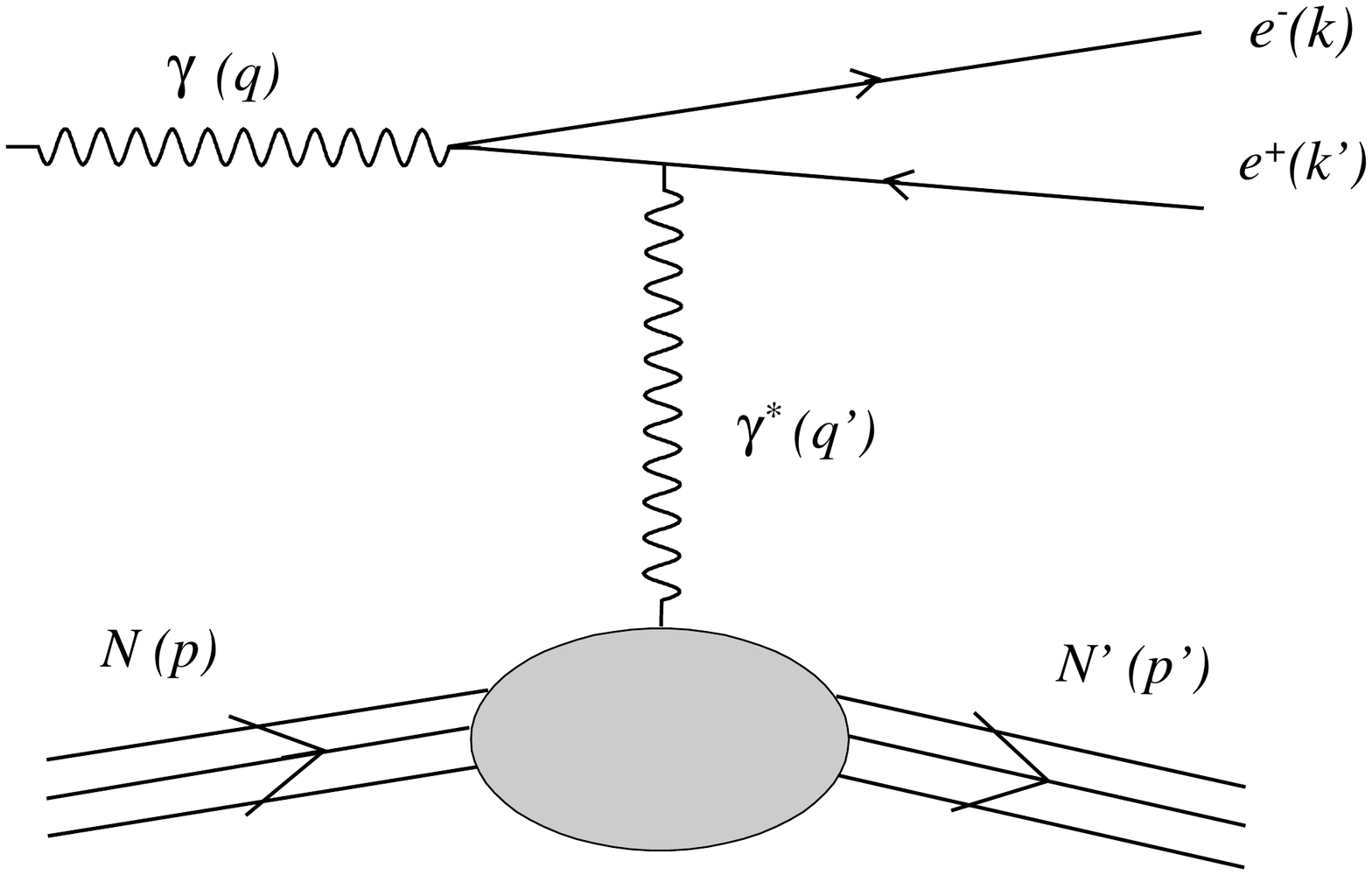}\\ \vspace*{-0.8cm}
\caption{Top: The direct TCS diagram (at QCD leading-twist); 
there is also a crossed diagram. Bottom: The BH 
diagram; there is also a diagram where the spacelike virtual
photon couples to the electron instead of the positron.}
\label{fig:TCSdiag}
\end{center}
\end{figure}

The $\gamma N\to N' e^+e^-$ 
reaction (where $N$ stands for a proton $p$ or a neutron $n$) consists of two processes: 
the Bethe-Heitler (BH) and the Timelike Compton Process (or TCS for Timelike Compton Scattering). See Fig.~\ref{fig:TCSdiag} for a sketch of the two processes. 
In the BH process, the final state lepton pair originates 
directly from the initial photon of the beam which is a pure QED (Quantum Electro-Dynamics) process. One of 
the leptons interacts then with the nucleon through the exchange
of a virtual spacelike photon. This process involves the elastic form factors of 
the nucleon. These are rather accurately known in the low nucleon momentum transfer 
$t$ region that concerns us.
The whole process is therefore quite precisely calculable. In the TCS process, the final state lepton 
pair originates from the timelike virtual photon which, if its virtuality $Q'^2$
is high enough,
is emitted off a quark from the target nucleon. Therefore, the final lepton pair kinematic 
distribution can be expected to reflect some aspects of the intrinsic dynamics
and interactions of the quarks in the nucleon. It can be shown in Quantum Chromo-Dynamics (QCD)
that, at low $t$ and large $Q'^2$, there is a factorization between the hard elementary Compton 
scattering at the quark level and the universal structure functions called 
GPDs~\cite{Mueller:1998fv,Rady96a,Ji97a,Ji97b}. These functions encompass 
the complex quark and gluon structure of the nucleon, which cannot be calculated at this time from the first
principles of QCD.

There are, at QCD leading twist, 4 chiral even nucleon GPDs entering the TCS process, : 
$H$, $E$, $\tilde{H}$ and $\tilde{E}$. They reflect the four independent quark helicity-nucleon spin
transitions between the initial and final states. Neglecting QCD
evolution effects, which is the farmework of this work, the GPDs depend on three kinematic variables 
$x, \xi, t$. In a frame where the nucleon approaches the speed of light, 
$x$ and $\xi$ define the initial and final quark longitudinal momentum fractions ($x+\xi$
and $x-\xi$ respectively, see Fig.~\ref{fig:TCSdiag}). 
The GPDs can then be interpreted as the probability amplitude to find
a quark in the initial (final) nucleon with a longitudinal momentum fraction
$x+\xi$ ($x-\xi$). In addition, the process involves a small transverse momentum transfer
which is contained in $t$. At $\xi=0$, GPDs provide access to the 
probability amplitude to find a quark in the nucleon with a
{\it longitudinal} momentum fraction $x$ at a given {\it transverse} impact 
parameter $b_\perp$, which is the conjugate variable of $t$.

This article is organized as follows. After a brief review of the formalism
in section II, we show the results of our calculations for different
observables in section III and present our conclusions in section III.

\section{Formalism}

We presented in Ref.~\cite{Boer:2015hma} the formalism that we used to
derive the amplitudes of the BH and TCS processes on the proton. 
We now make a brief recap, which we adapt to the neutron target case. 

%We define the average neutron momentum $P$, neutron momentum transfer $\Delta$
%and average photon momentum $\bar{q}$:

%\begin{eqnarray}
%&&P = {1 \over 2} \left(p + p' \right), \\
%&&\Delta=(p'-p)=(q-q'),
%&&\bar{q} = {1 \over 2} \left(q + q' \right) .
%\end{eqnarray}

%The final photon virtuality is related to the average photon momentum as: 

%\begin{equation}
%$\bar{q}^2=Q'^2 /2 -\Delta^2/4$
%\end{equation}

We use a frame where the average photons momenta $\bar{q}={1 \over 2} \left(q + q' \right)$ 
and neutrons momenta $N={1 \over 2} \left(p + p' \right)$ are collinear along the $z$-axis 
and in opposite directions. We define the lightlike vectors along the positive and negative 
$z$ directions as:  
\begin{eqnarray}
&&\tilde p^\mu = P^+/\sqrt{2} (1,0,0,1), \\  
&&n^\mu = 1/\sqrt{2}P^+ (1,0,0,-1),
\end{eqnarray}
where the light-cone components $a^\pm$ are defined by $a^\pm\equiv (a^0 \pm a^3)/\sqrt{2}$.

In this frame, using Ji's conventions for the GPDs~\cite{Ji97a,Ji97b}, the TCS amplitude reads:
\begin{eqnarray}
\label{eq:T_TCS}
&&T^{TCS} = -\frac{e^3}{q'^2} \:\bar{u}(k)\:\gamma^{\nu} \: v(k')\:\epsilon^{\mu}(q)\: \\ \nonumber
&&\times\Bigg\{\frac{1}{2}\:(-g_{\mu\nu})_{\perp}
\int\limits_{-1}^{1}dx\: \left(
\frac{1}{x-\xi-i\epsilon} + \frac{1}{x+\xi + i\epsilon}\right) \\ \nonumber
&&\times\: \bigg(
H^n(x,\xi,t) \bar{u}(p')\slashed{n}u(p) \\ \nonumber
&&+E^n(x,\xi,t)\bar{u}(p') i \sigma^{\alpha\beta}n_{\alpha} \frac{\Delta_{\beta}}{2m}\:u(p)\bigg) \\ \nonumber
&&-\frac{i}{2}(\epsilon_{\nu\mu})_{\perp} 
\int\limits_{-1}^{1}dx\: \bigg(
\frac{1}{x-\xi-i\epsilon} - \frac{1}{x+\xi + i\epsilon}
\bigg)\\ \nonumber
&&\times\bigg(\tilde{H}^n(x,\xi,t)\bar{u}(p')\slashed{n}\gamma_5 \:u(p) \\ \nonumber
&&+ \tilde{E}^n(x,\xi,t)\bar{u}(p')\gamma_5 \frac{\Delta.n}{2m}\:u(p)
\bigg)\Bigg\},
\end{eqnarray}
where we have used the metrics: 
\begin{eqnarray}
&&(-g_{\mu\nu})_{\perp}= -g_{\mu\nu} + \tilde{p}_{\mu}n_{\nu}   + \tilde{p}_{\nu} n_{\mu} \:\:,\\ \nonumber
&&(\epsilon_{\nu\mu})_{\perp}=\epsilon_{\nu\mu\alpha\beta}\: n^{\alpha}\: \tilde{p}^{\beta}.
\end{eqnarray}

In Eq.~\ref{eq:T_TCS}, $m$ is the neutron mass and 
the light-cone momentum fractions $x$ and $\xi$ are defined respectively by $k^+=xP^+$, 
and by $\Delta^+=-2\xi P^+$ where $\Delta^\mu=(p'-p)^\mu=(q-q')^\mu$.
We have $\xi =$${Q'^2}\over{2(s-m^2)-Q'^2}$, where $s=(p+q)^2$,
when we neglect $\Delta$ terms 
w.r.t. to $Q'^2$. In the following, we will place ourselves in this limit.
In Ref.~\cite{Boer:2015hma}, we compared this limit with the exact kinematics and
formulaes and we found that the effects on the observables associated to 
this approximation were negligible for the Jefferson Lab (JLab) kinematics discussed in the following.
Regarding the GPDs, for numerical estimates, we will use the parameterization given
by the VGG 
model~\cite{Vanderhaeghen:1998uc,Vanderhaeghen:1999xj,Goeke:2001tz,Guidal:2004nd}.
The $x,\xi$-dependence follows the double-distribution ansatz~\cite{Radyushkin:1998es,Radyushkin:1998bz,
Mueller:1998fv} and the $t$-dependence is  
Reggeized~\cite{Goeke:2001tz,Guidal:2004nd}.
The VGG model contains also the so-called D-term~\cite{Polyakov:1999gs}
whose influence will be studied in the following. 

GPDs are defined for each quark flavor. Restricting ourselves to the $u$ and $d$
flavors, we have the following decomposition for the neutron and the proton:

\begin{eqnarray}
&&GPD^n(x,\xi,t) \,=\,{1 \over 9} GPD^{u} \,+\, {4 \over 9} GPD^{d},\\ \nonumber
&&GPD^p(x,\xi,t) \,=\,{4 \over 9} GPD^{u} \,+\, {1 \over 9} GPD^{d}.
\end{eqnarray}

The simultaneous study of TCS off the proton and off the neutron therefore allows 
to make a flavor separation of GPDs.

The BH amplitude reads:
\begin{eqnarray}
&&T^{BH} = -\frac{e^3}{\Delta^2} \:\bar{u}(p')\: \Gamma^{\nu}\: u(p)\: \epsilon^{\mu}(q) \\ \nonumber
&& \times\bar{u}(k)\left(
\gamma_{\mu}   
\frac{\slashed{k}-\slashed{q}}{(k-q)^2}
\gamma_{\nu}
+
\gamma_{\nu}   
\frac{\slashed{q}-\slashed{k'}}{(q-k')^2}
\gamma_{\mu}
\right)
\: v(k'), 
\end{eqnarray}
with the virtual photon-neutron electromagnetic vertex matrix  
\begin{equation}
\Gamma^{\nu} = \gamma^{\nu} \:F_1^n(t)  + \frac{i \sigma^{\nu\rho} \Delta_{\rho} }{2\:m}\:F_2^n(t). \:
\end{equation}
$F_1^n(t)$ and $F_2^n(t)$ are the neutron Dirac and Pauli form 
factors. In this work, we take the parametrizations
based on Refs.~\cite{Warren:2003ma,Kubon:2001rj}.

At fixed beam energy $E_\gamma$ or longitudinal 
momentum transfer fraction $\xi$, there are four independent 
kinematical variables for the process
$\gamma N\to N' e^+e^-$. We choose them
as: $Q'^2$ and $t$ ($=\Delta^2$) that we already defined,  
and the two angles $\theta$ and $\phi$ of the electron in the 
leptons' center-of-mass frame. The polar angle $\theta$ is defined 
w.r.t. to the $z'$-axis which correspond to the direction of the two-lepton system 
in the $(e^+e^-)-N'$ center of mass and $\phi$ is the azimutal angle between
the decay plane and the production plane. We refer the reader to 
Fig.4 of Ref.~\cite{Boer:2015hma} for an illustration and the orientation of the angles.

\section{Calculation of observables}
\subsection{Unpolarized cross section}

The 4-fold differential unpolarized cross section is expressed as:
\begin{eqnarray}
\frac{d^4\sigma}{dQ'^2dtd\Omega }({\gamma N \to Ne^+e^-}) 
= &&\frac{1}{2\pi^4}\frac{1}{64}\frac{1}{2mE_\gamma}\\ \nonumber
&&\times \mid T^{BH} + T^{TCS}\mid^2
\end{eqnarray}
where 
$T^{BH} + T^{TCS}$ is summed over the final nucleon and final electron helicities and
is averaged over the target nucleon helicities and beam polarizations. 

\begin{figure}[htbp]
\begin{center} 
\includegraphics[width=11.cm]
{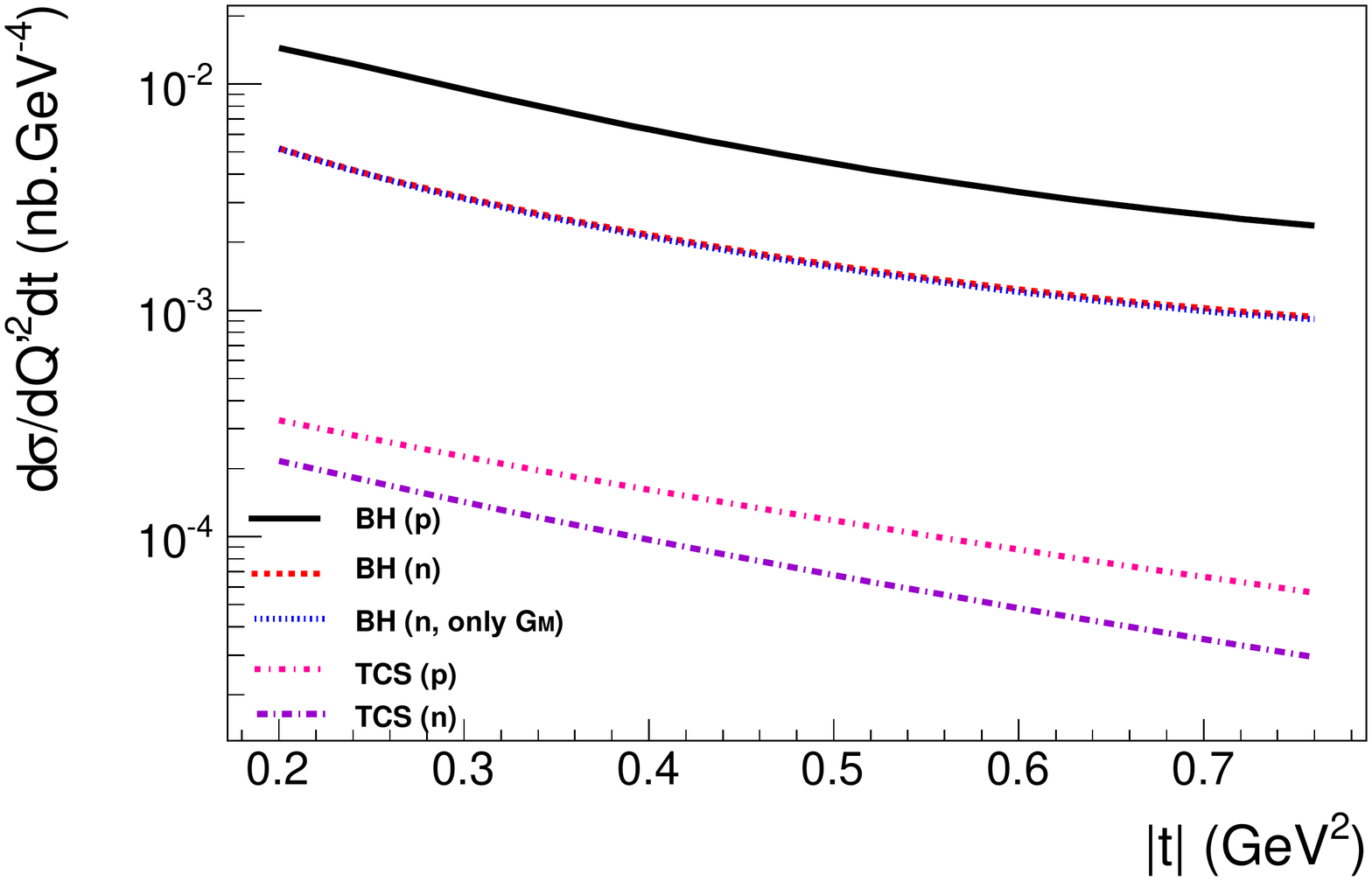}
\vspace*{-0.5cm}
\caption{Unpolarized cross sections $\frac{d\sigma_{BH}}{dQ'^2\:dt}$
for BH and for TCS off the neutron and proton. The calculations have
been done at $\xi=0.2$ and $Q'^2=7$ GeV$^2$. The angles $\theta$ and 
$\phi$ have been integrated 
over $[\pi/4, 3\pi/4]$ and $[0,2\pi]$ respectively.
The neutron BH calculation with only the neutron magnetic form factor
contribution is also shown.}
\label{fig:unpol}
\end{center}
\end{figure}

We present in Fig.~\ref{fig:unpol} the results of our calculations for
the $t$-dependence of the two-fold unpolarized TCS and BH cross sections 
$\frac{d\sigma}{dQ'^2\:dt}$. The calculations have
been performed at $\xi=0.2$ and $Q'^2=7$ GeV$^2$, which is a typical 
kinematical setting accessed at JLab 12 GeV. The angles $\theta$ and $\phi$ have been integrated 
over $[\pi/4,3\pi/4]$ and $[0,2\pi]$ respectively.
The general motivation for integrating over $\theta$, which we will do
for all observables in the following, is to maximize the count rates 
from an experimental point of view. We showed in Ref.~\cite{Boer:2015hma} 
that this did not result in an important loss of sensitivity to TCS,
and therefore to GPDs, compared to a fixed $\theta$ value.
The reason for limiting the range of integration in $\theta$ is that the BH
exhibits (quasi-)singularities around $\theta$=0$^\circ$ ($\theta$=180$^\circ$).
These regions correspond to cases where the electron 
(positron) is emitted in the direction of the initial photon, so that the
lepton propagator becomes singular and creates a sharp peak in the cross section.
In order to optimize the sensitivy to TCS, it is therefore advisable
to keep away from the BH-dominated regions such as $\theta$=0$^\circ$ ($\theta$=180$^\circ$).

In Fig.~\ref{fig:unpol}, we compare $\frac{d\sigma}{dQ'^2\:dt}$ 
for the neutron and proton target cases. 
For both, the BH cross section is more than one order of magnitude 
higher than the TCS one. As a consequence, as we discussed in Ref.~\cite{Boer:2015hma},
the sensitivity to TCS, and thus to GPDs, will better show up through
interference effects, and therefore spin observables.
Both the BH and TCS neutron cross sections are less than a factor of two lower
than the corresponding proton cross sections. The neutron channel is therefore
in principle measurable experimentally, modulo
neutron detection efficiency issues. We recall
that a proposal for measuring TCS on the proton at JLab has already been
approved~\cite{JLABprop}. We also display in Fig.~\ref{fig:unpol}
the neutron BH cross section calculated with only the neutron magnetic form factor
contribution. This curve is almost indistinguishible from the calculation
with both form factors, which shows that the neutron BH cross section 
is largely dominated by the contribution of the neutron magnetic form factor.

\vspace*{-0.5cm}

\subsection{Single spin asymmetries}

In this section, as well as in the following one on double spin asymmetries,
our calculations are carried out for 100\% polarisation.
Circularly polarized photons, linearly polarized photons and polarized targets
with high degrees of polarization (between 60\% and 90\%) have been obtained and 
used these past years almost routinely at JLab (see for 
instance Ref.~\cite{Pasyuk:2009yt}).

Following the conventions of Ref.~\cite{Boer:2015hma}, we use the notation $A_{ij}$
for the spin asymmetries where the first index refers to the polarization
of the beam and the second one to the polarization of the target. There are 
three sorts of single-spin asymmetries: 

\begin{eqnarray}
&&A_{\odot U} = \frac{\sigma^{+} - \sigma^{-}}{\sigma^{+} + \sigma^{-}},\\ 
&&A_{\ell U}(\Psi) = \frac{\sigma(\Psi) - \sigma(\Psi+\pi/2)}
{\sigma(\Psi) + \sigma(\Psi+\pi/2)}, \\
&&A_{Ui} = \frac{\sigma^{+} - \sigma^{-}}{\sigma^{+} + \sigma^{-}}. 
\label{eq:defass}
\end{eqnarray}

They are associated respectively to a circulary polarized beam, a linearly polarized beam 
and a polarized target. The index $U$ stands for ``unpolarized".
For the target spin asymmetries $A_{Ui}$, the index $i=x,y,z$ corresponds to the polarization 
direction of the target 
along the $x$, $y$, $z$ axis respectively. The $z$-axis is along the photon direction 
in the
$\gamma-N$ center of mass system, the $y$-axis is perpendicular to the reaction plane
and the $x$-axis is perpendicular to the target nucleon direction and in the reaction 
plane (see Fig.4 of Ref.~\cite{Boer:2015hma} for an illustration of our axis 
orientation convention).
The superscripts $\pm$ refer to the target spin orientation along those axis.
For the circularly polarized asymmetry $A_{\ell U}$, $\Psi$ is the angle between 
the polarization vector of the incoming photon and the scattering plane.
For the linearly polarized asymmetry $A_{\odot U}$, the $+$ superscript
refers to the right circular polarization and the $-$ superscript
to the left circular polarization.

\begin{figure}[htbp]
\begin{center} 
\vspace*{-1cm}\hspace*{-0.5cm}
\includegraphics[width=9cm,height=7.5cm]
{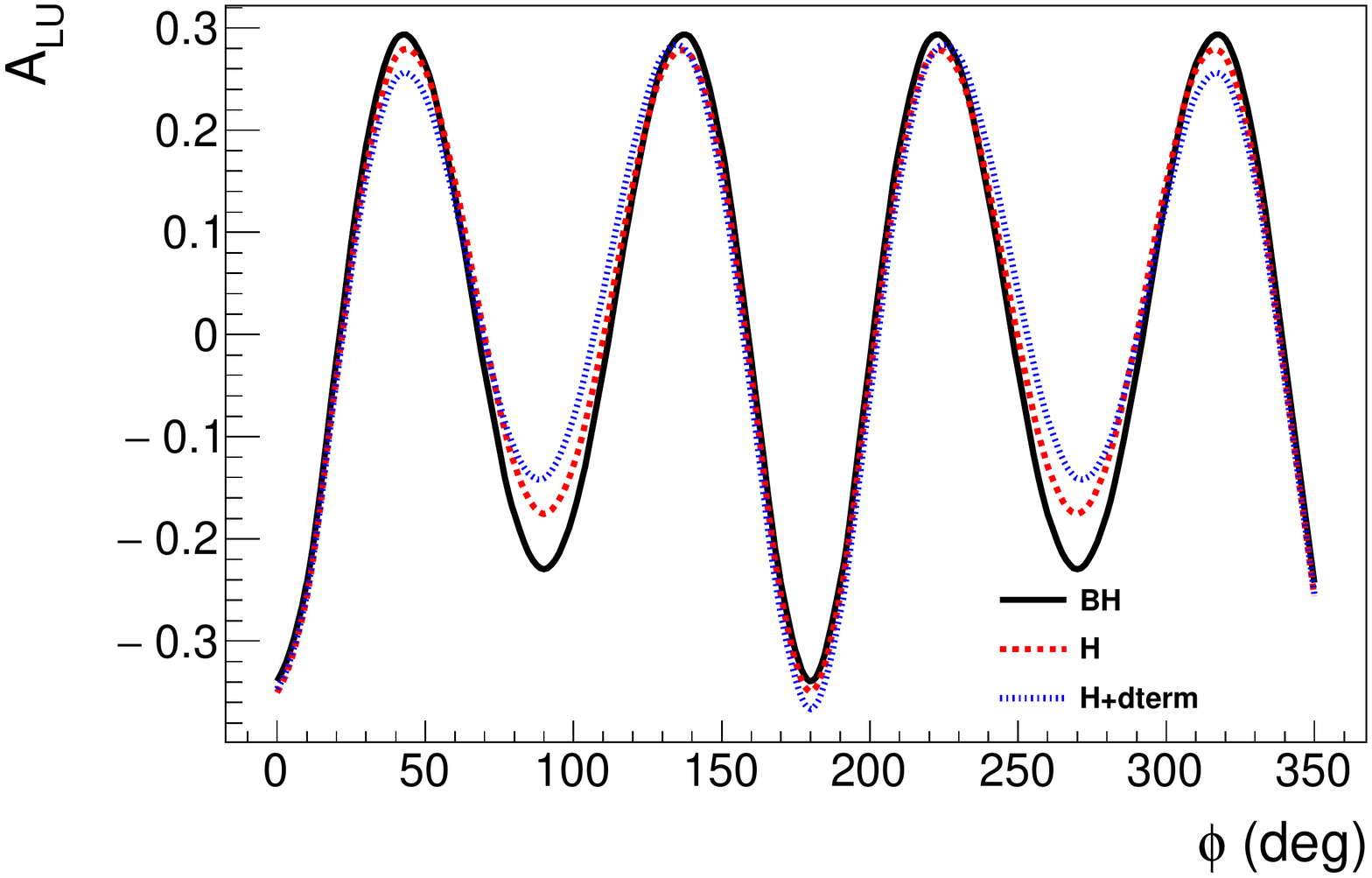}
\includegraphics[width=9cm,height=7.5cm]
{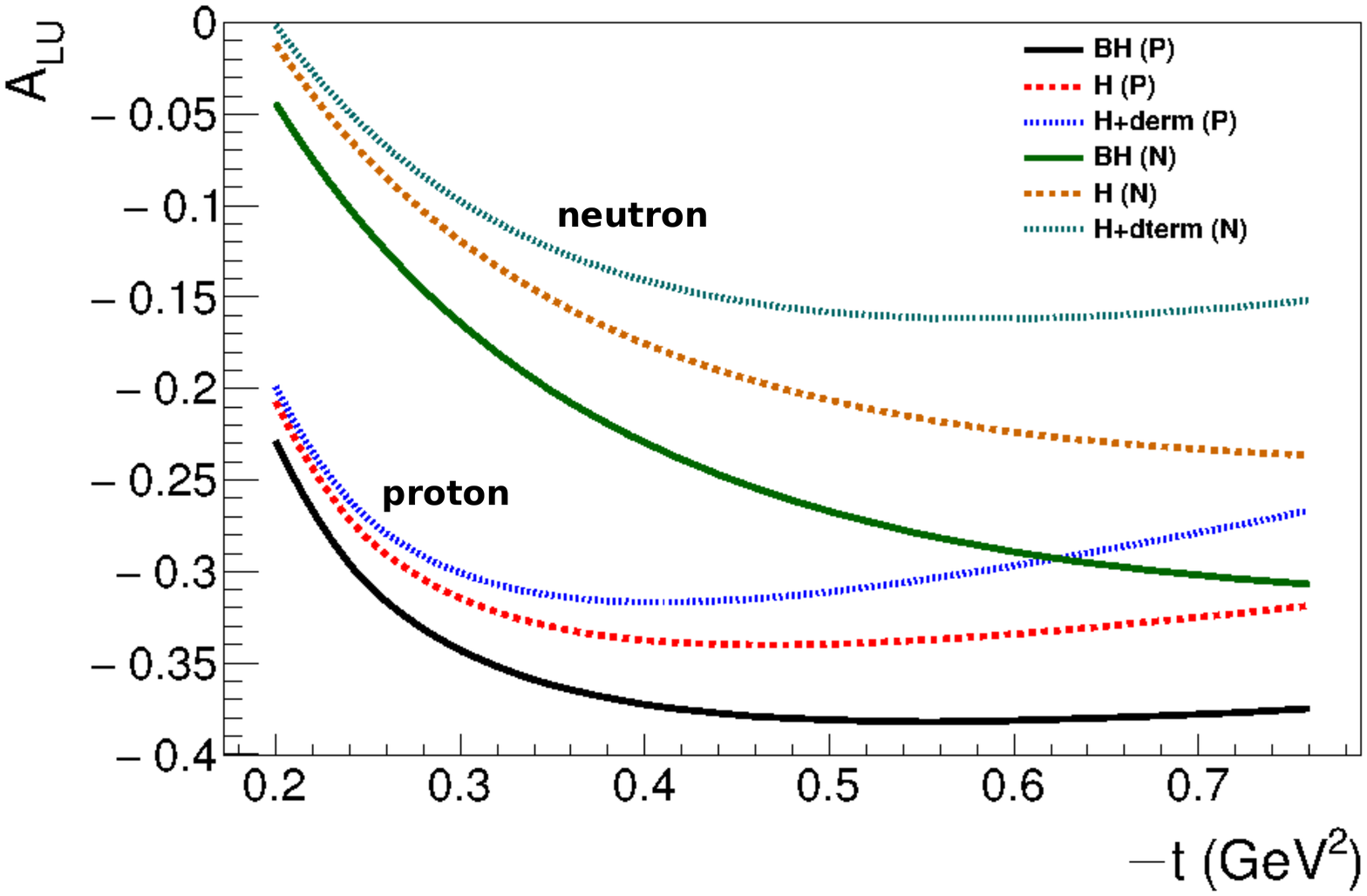}
\\\vspace*{-1cm}
\caption{Top panel: $A_{\ell U}$
as a function of $\phi$ at $\xi=0.2$, $Q'^2=7$ GeV$^2$, $-t=0.4$ GeV$^2$ and 
with $\theta$ integrated over $[\pi/4,3\pi/4]$. Bottom panel: $A_{\ell U}$
as a function of $t$ for $\phi=90^\circ$ at $\xi=0.2$, $Q'^2=7$ GeV$^2$ 
and $\theta$ integrated over $[\pi/4,3\pi/4]$.}
\label{fig:BSAl}
\end{center}
\end{figure}

We begin by studying the linearly polarized beam single-spin asymmetry.
Similarly to the proton target case, $A_{\ell U}$ shows
a typical $\cos(2\Psi)$-like modulation which we do not show here.
In the following, we therefore calculate $A_{\ell U}$ for $\Psi=0^\circ$.
The upper panel of Fig.~\ref{fig:BSAl} shows the $\phi$-dependence of $A_{\ell U}$
for the $\gamma n\to n' e^+e^-$ reaction 
for $\Psi=0^\circ$, $\xi=0.2$, $Q'^2=7$ GeV$^2$, $-t=0.4$ GeV$^2$ and $\theta$ integrated 
over $[\pi/4,3\pi/4]$.
This observable is sensitive to the real part of the BH + TCS amplitude. 
We recall that the BH amplitude is purely
real. It is seen in Fig.~\ref{fig:BSAl} that the BH alone produces actually most 
of the signal, with an amplitude around 30$\%$. The TCS, essentially
through the GPD $H^n$, brings only small variations around $\phi=90^\circ$ 
w.r.t. the BH signal. The introduction of the $\tilde H^n$, $E^n$ and
$\tilde E^n$ GPDs barely changes the asymmetry calculated with only $H$.
However, the introduction of the D-term is noticeable. 
 In the bottom panel of Fig.~\ref{fig:BSAl},
we show the $t$-dependence of $A_{\ell U}$ for $\phi=90^\circ$ for both the neutron and proton cases
(this latter asymmetry was calculated in Ref.~\cite{Boer:2015hma}).
The amplitude of both asymmetries increases with $\mid t\mid$, faster for the neutron
than for the proton. In both cases, the TCS contribution diminishes the amplitude of the asymmetry. 
The sensitivity of the asymmetry to the GPDs grows with increasing values of $\mid t\mid$, 
with a dominating influence of $H^n$ and of the D-term.

\begin{center}
\begin{figure}[htbp]
\hspace*{-0.8cm}
\includegraphics[width=8cm,height=6cm]
{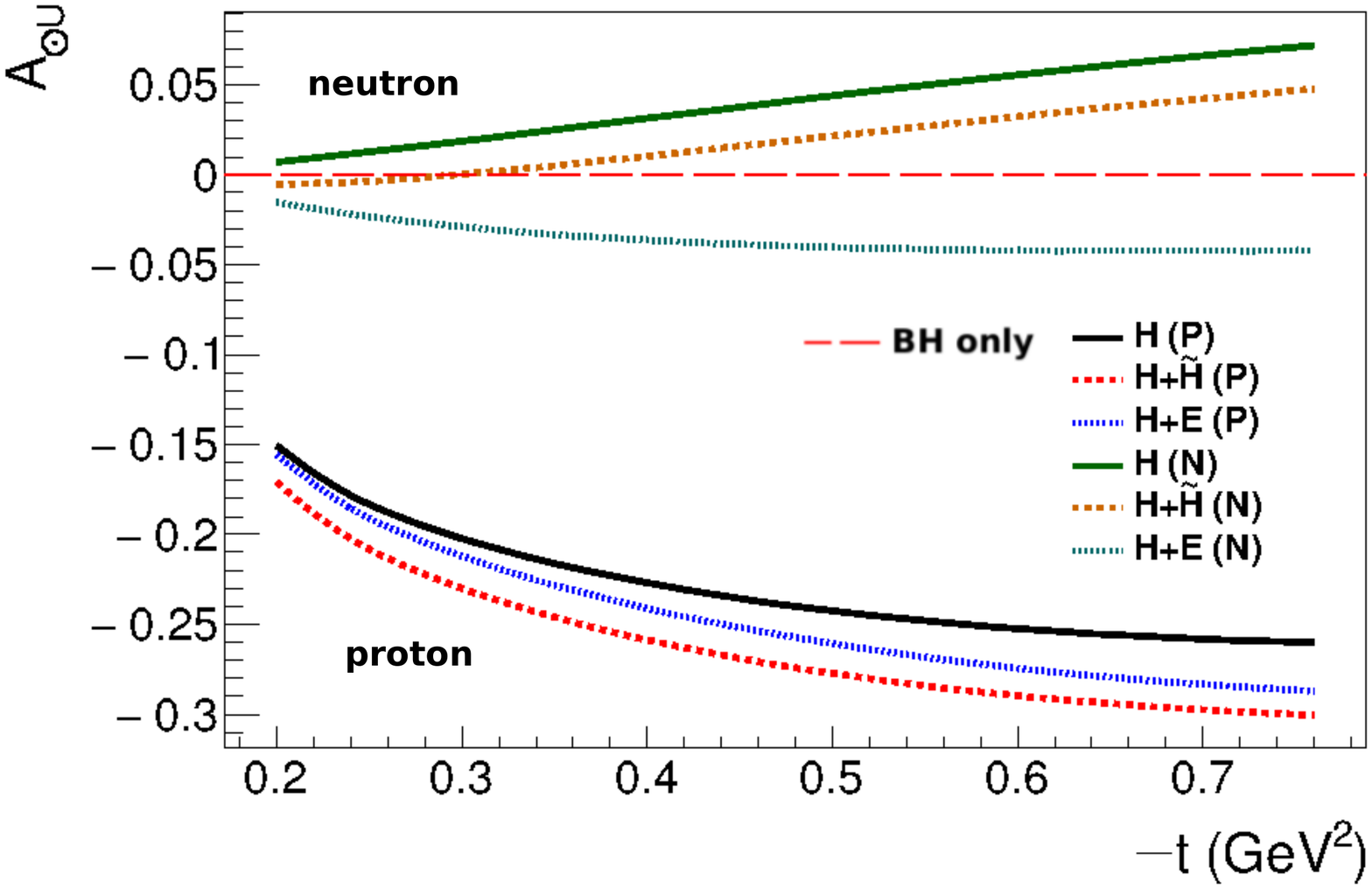}
\caption{$A_{\odot U}$, for the neutron and proton target cases,
as a function of $t$ for $\phi=90^\circ$ at $\xi=0.2$, $Q'^2=7$ GeV$^2$ and $\theta$ integrated 
over $[\pi/4,3\pi/4]$.}
\label{fig:BSAc}
\end{figure}
\end{center}

In Fig.~\ref{fig:BSAc}, we show the other beam single-spin asymmetry $A_{\odot U}$, 
i.e. obtained with a circularly polarized photon beam. Like for the proton case~\cite{Boer:2015hma}, the
asymmetry exhibits for the $\gamma n\to n' e^+e^-$ reaction
a $\sin\phi$-like shape and we display in Fig.~\ref{fig:BSAc} $A_{\odot U}$
as a function of $t$ for $\phi=90^\circ$ (at $\xi=0.2$, $Q'^2=7$ GeV$^2$ and $\theta$ integrated 
over $[\pi/4,3\pi/4]$). One notices that the BH alone produces a zero asymmetry. This is due to
the fact that the $A_{\odot U}$ observable is sensitive to the imaginary
part of the amplitude, making this observable very favorable for the
study of TCS and GPDs. When the TCS process is included, the asymmetries on the neutron
target reach up to 5\% at the largest values of $\mid t\mid$ considered here. 
From Fig.~\ref{fig:BSAc}, one sees that this observable has a sensitivity
to $H^n$ but also to $\tilde H^n$ and $E^n$. It is particularly 
interesting to note that when 
only $H^n$ is taken into account, $A_{\odot U}$ for the neutron case is positive
while it becomes negative when $E^n$ is introduced. The explanation
is the following. The asymmetry results from the
interference of the BH and TCS processes. It is therefore proportional to the product of
the two amplitudes. As we noticed earlier, the neutron BH is dominated by the nucleon magnetic 
form factor contribution which has an opposite sign to the proton electric and magnetic form factors.
On the TCS side, $H^n$ has the same sign than $H^p$. The product of the BH and
of the TCS processes, and therefore the asymmetry, has thus an opposite
sign between the neutron and proton cases when only $H^n$ is taken into account. In contrast, $E^n$
has an opposite sign to $E^p$ and therefore the asymmetry becomes, like for the proton, negative
when the GPD $E^n$ is included. This asymmetry appears therefore rather interesting to study
the GPD $E^n$.

We recall that the GPD $E$ is one of the two GPDs entering Ji's sum rule:

\begin{equation}
J_q \,=\, {1 \over 2} \, \int_{-1}^{+1} d x \, x \, 
\left[ H^{q}(x,\xi,t = 0) + E^{q}(x,\xi,t = 0) \right] \;, 
\label{eq:jisr}
\end{equation}
linking the total angular momentum ($J_q$) carried by quarks of
flavor $q$ to the sum of the second moments over $x$ of the GPDs $H$ and $E$.
The GPD $H$ can be considered as relatively well constrained due to its
model-independent relations with parton densities and form factors.
This is much less the case for the GPD $E$ whose forward limit is not
constrained at all. The unknown part of $J_q$ therefore lies essentially
in the GPD $E$. In the VGG model, the GPD $E^q$ is parametrized 
as a function of $J_q$ which is taken as
a free parameter~\cite{Goeke:2001tz}. The idea is that the VGG model assumes a certain shape in $x$ 
for the GPD $E^q$ and then the overall normalization of this function is proportional to
$J^q$. We refer to Ref.~\cite{Goeke:2001tz} for details. Although this relation 
between $E^q$ and $J^q$ is clearly model-dependent, it yields estimates
for $J^u$ and $J^d$ which are in agreement with other approaches, such as
lattice QCD~\cite{Alexandrou:2014yha}.
 
\onecolumngrid

\begin{center}
\begin{figure}[htbp]
\hspace*{-0.8cm}
%\includegraphics[width=8cm,height=8cm]
%{LegoDistr_EvolJuJd_A0U_TCSneutr.pdf}
%\hspace*{-0.8cm}
%\includegraphics[width=8cm,height=8cm]
%{LegoDistr_EvolJuJd_ALU_DVCSneutrii.pdf}
%\hspace*{-0.8cm}
%\includegraphics[width=8cm,height=8cm]
%{LegoDistr_EvolJuJd_A0U_TCSprot.pdf}
%\hspace*{-0.8cm}
%\includegraphics[width=8cm,height=8cm]
%{LegoDistr_EvolJuJd_ALU_DVCSprotii.pdf}
\includegraphics[width=15cm,height=15cm]
{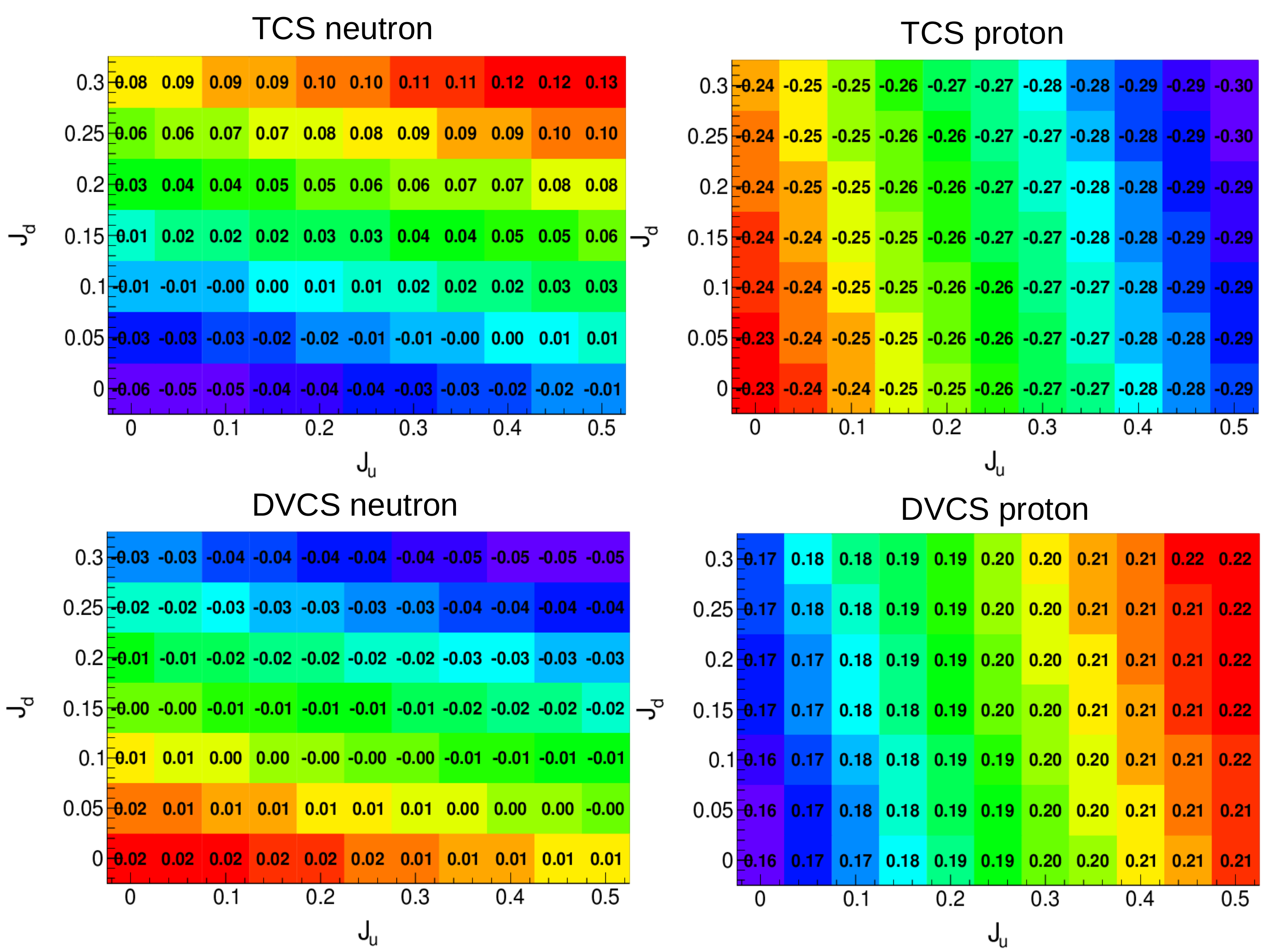}
\hspace*{-0.8cm}
\caption{TCS $A_{\odot U}$ on the neutron (top left panel), 
TCS $A_{\odot U}$ on the proton (top right panel), 
DVCS $A_{LU}$ on the neutron (bottom left panel), 
DVCS $A_{LU}$ on the proton (bottom right panel) as a function
of $J^u$ and $J^d$.
Calculations are done for $\phi=90^\circ$, $\xi=0.2$, 
$Q'^2=7$ GeV$^2$, $-t=0.4$ GeV$^2$ and $\theta$ integrated 
over $[\pi/4,3\pi/4]$.}
\label{fig:jujd}
\end{figure}
\end{center}

\twocolumngrid

Since $A_{\odot U}$ on the neutron appears to be mainly governed by the GPD $E$,
it can be interesting, in the model-dependent framework of VGG that we just described, 
to calculate $A_{\odot U}$ directly as a function of $J^u$ and $J^d$.
We show in the top letf panel of Fig.~\ref{fig:jujd} the results
of our calculations for $A_{\odot U}$ at $\phi=90^\circ$, $\xi=0.2$, 
$Q'^2=7$ GeV$^2$, $-t=0.4$ GeV$^2$ and $\theta$ integrated 
over $[\pi/4,3\pi/4]$) for different values of $J^u$ and $J^d$.
For the total angular momentum values considered, the figure shows 
the rather strong sensitivity of the asymmetry 
which varies from -6\% to 13\%. In the bottom left panel of Fig.~\ref{fig:jujd},
for comparison, we show, in the same VGG framework, the sensitivity of the 
neutron DVCS beam spin asymmetry $A_{LU}$ to $J^u$ and $J^d$. One notices as well
the sensitivity of this observable to $J^u$ and $J^d$ with in particular
also changes of sign. However, the amplitudes of the asymmetries for the neutron DVCS
case are about a factor of 3 smaller than in the neutron TCS case. 
We recall that there is an approved experimental proposal with the
CLAS detector at JLab to measure the neutron DVCS beam spin asymmetry~\cite{nDVCS}.
Although count rates for TCS can be expected to be less important
than for DVCS, the larger TCS asymmetry can possibly provide an interesting
alternative way to access the GPD $E$.

As a side remark, we note that TCS and DVCS asymmetries have opposite
signs w.r.t. each other, i.e. the TCS asymmetry is positive
when the DVCS asymmetry is negative and vice-versa. It was 
indeed shown in Ref.~\cite{Muller:2012yq} that the TCS amplitude is the conjugate 
of the DVCS amplitude. The TCS $A_{\odot U}$ and DVCS $A_{LU}$ asymmetries
being proportional to the imaginary part of the TCS+BH amplitude, this
naturally explains this opposite sign.

For completeness, we also show in Fig.~\ref{fig:jujd} the sensitivity
of the proton TCS and DVCS beam spin asymmetries to $J^u$ and $J^d$. One
sees that there is no change of sign of these asymmetries as a function $J^u$ and $J^d$
in contrast to the neutron case. This is because the proton beam spin
asymmetries are dominated by the GPD $H$. We also remark the relative
opposite sign of the TCS and DVCS asymmetries w.r.t. each other, as for the neutron.

\begin{center}
\begin{figure}
\hspace*{-0.8cm}
\includegraphics[width=6.5cm,height=6cm]
{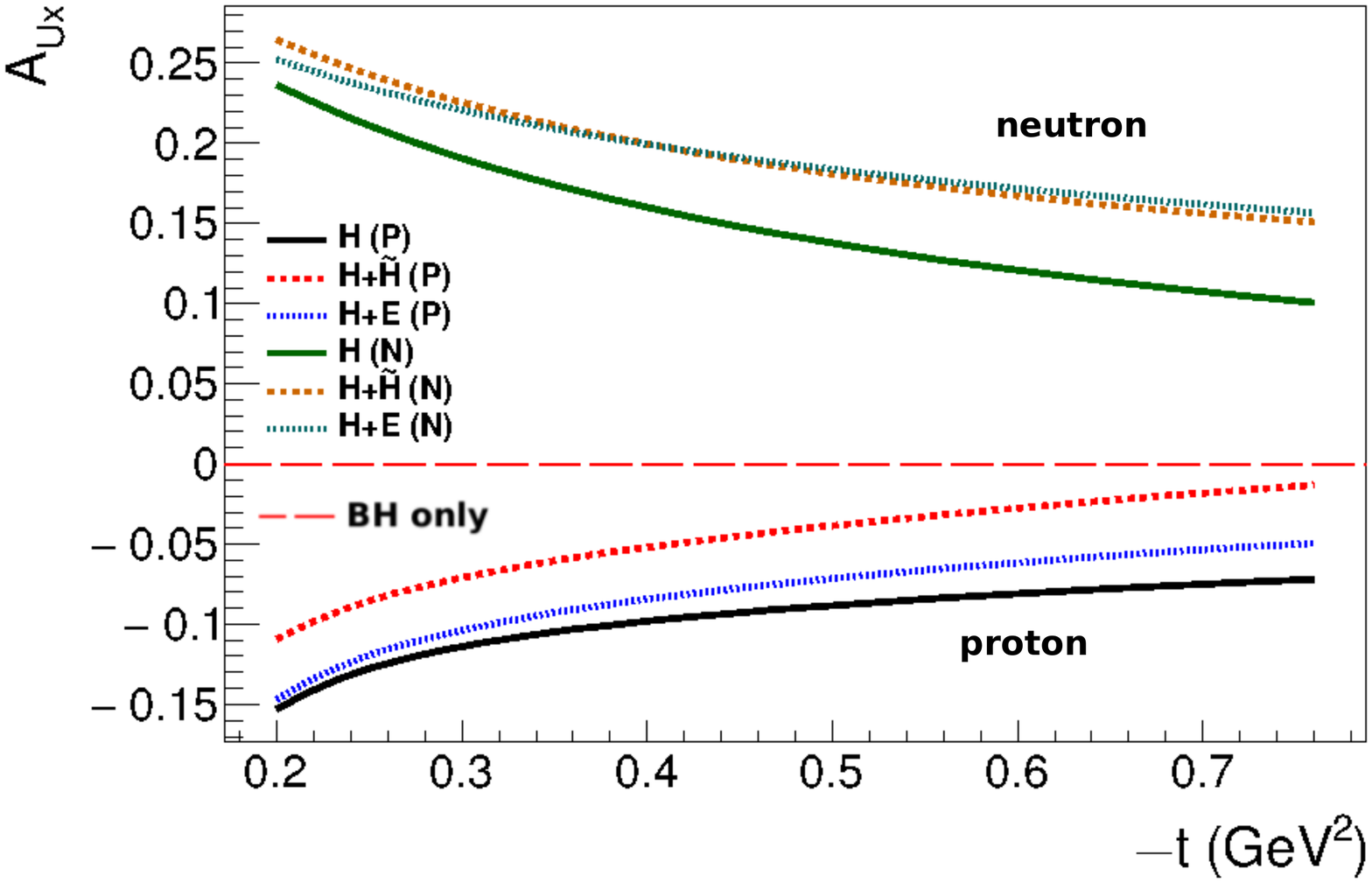}
\hspace*{-0.8cm}
\includegraphics[width=6.5cm,height=6cm]
{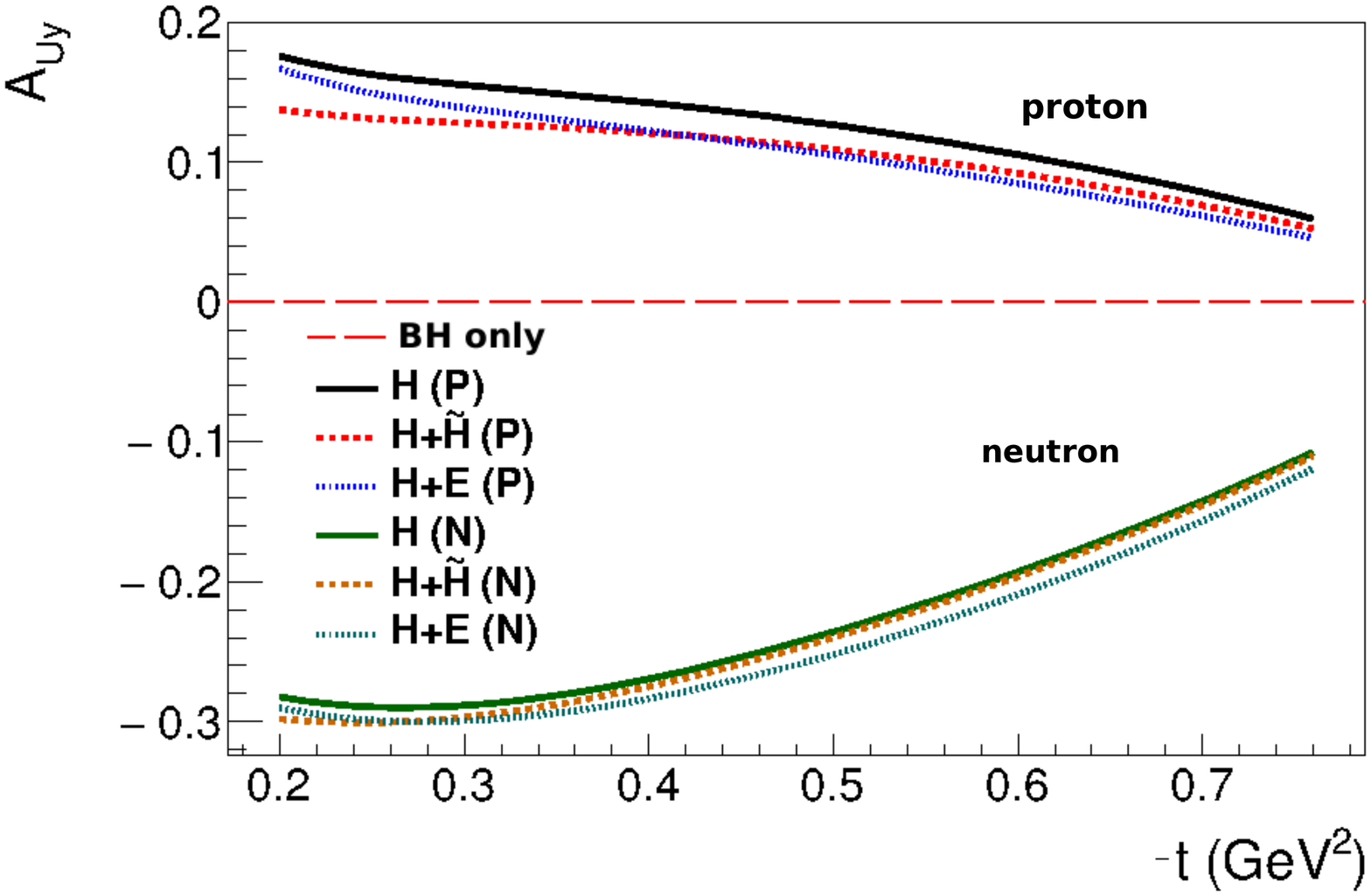}
\hspace*{-0.8cm}
\includegraphics[width=6.5cm,height=6cm]
{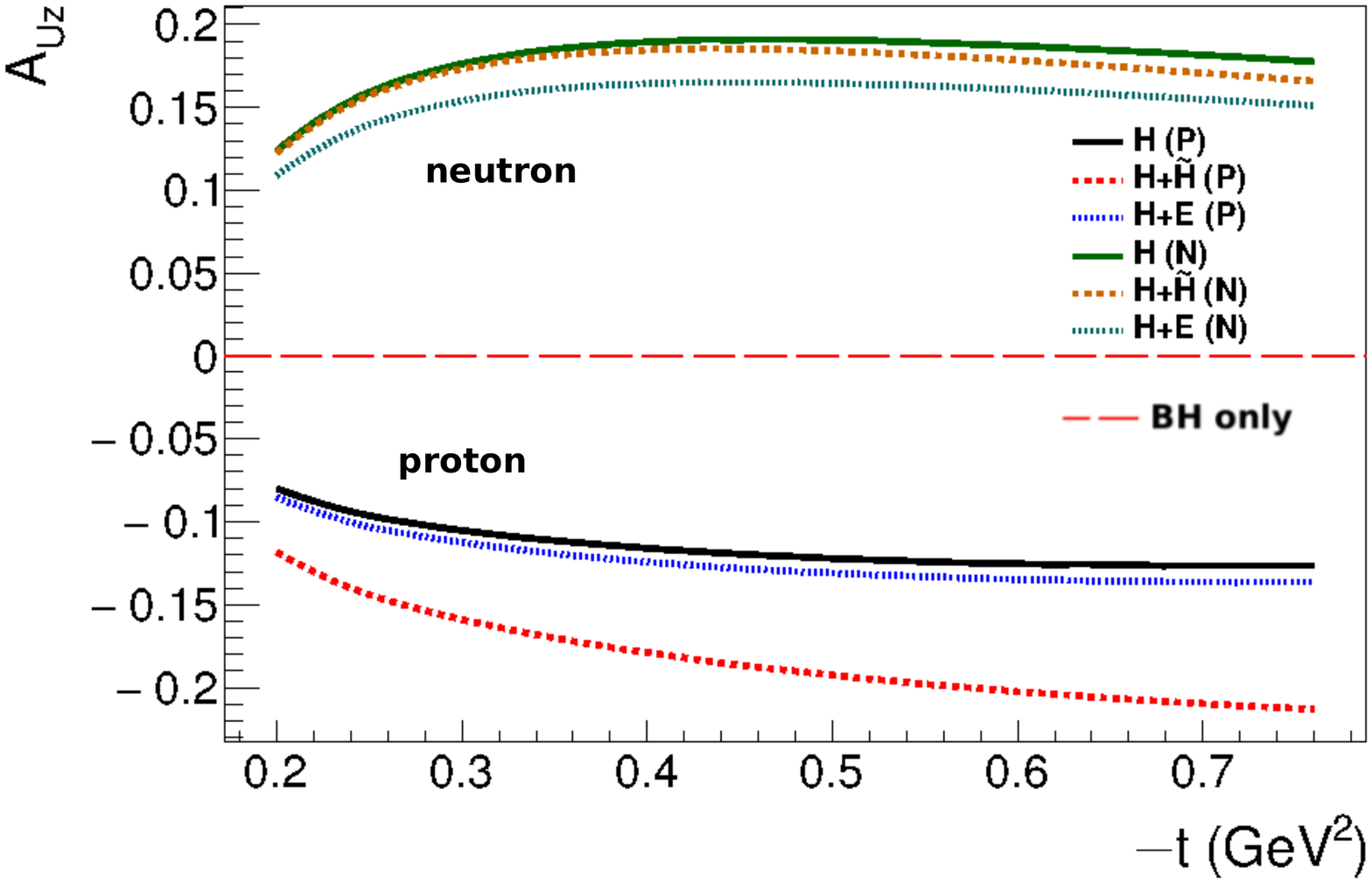}
\hspace*{-0.8cm}
\caption{The $A_{Ux}$ (left panel), $A_{Uy}$ (central panel) 
and $A_{Uz}$ (right panel)
asymmetries, as a function of $t$, respectively for
$\phi=90^\circ$, $\phi=0^\circ$ and $\phi=90^\circ$
at $\xi=0.2$, $Q'^2=7$ GeV$^2$,
$-t=0.4$ GeV$^2$ and for $\theta$ 
integrated over $[\pi/4,3\pi/4]$.
The results are shown for the neutron and proton target cases.}
\label{fig:TSA}
\end{figure}
\end{center}

We now turn to the target single-spin asymmetries. 
The asymmetries $A_{Ux}$, $A_{Uy}$ and $A_{Uz}$ have respectively
a $\sin\phi$, $\cos\phi$ and $\sin\phi$-like shapes. We therefore display in
Fig.~\ref{fig:TSA} these three asymmetries for, respectively, $\phi=90^\circ$,
$\phi=0^\circ$ and $\phi=90^\circ$, as a function of $-t$,
at the kinematics $\xi=0.2$, $Q'^2=7$ GeV$^2$ 
and $\theta$ integrated over $[\pi/4,3\pi/4]$. The figure shows both
the neutron and proton cases. 
Like for $A_{\odot U}$, these observables are sensitive to the imaginary part
of the amplitude and therefore the BH alone asymmetry is zero. 
We also observe an opposite sign for the two kinds of target when TCS is included
due to the different sign of the neutron magnetic form factor compared
to the proton. All $A_{Ui}$'s are
dominated by the GPD $H^n$. We however note that $A_{Ux}$ shows in addition a sensitivity 
to the GPDs $E^n$ and $\tilde H^n$ for the neutron case. Likewise, $A_{Uz}$ shows an additional 
sensitivity
to $E^n$ (while to $\tilde H^p$ for the proton case). 

\onecolumngrid

\twocolumngrid

\subsection{Double spin asymmetries}

We define the double-spin asymmetries as:
\begin{eqnarray}
&&A_{(\ell, \odot) i}= \frac{ (\sigma^{++} + \sigma^{--}) - ( \sigma^{+-} +  \sigma^{-+}) }
{\sigma^{++} + \sigma^{--}+\sigma^{+-} +  \sigma^{-+}},
\end{eqnarray}
where the first superscript $\pm$ refers to the polarization nature of the beam
($\ell$ for a linearly polarized photon beam and $\odot$
for a circularly polarized photon beam)
and the second one to the target's along the axis $i=x,y,z$.

%In Fig.~\ref{fig:BTSAlinphi}, we show the $\phi$-dependence of 
%the three double spin asymmetries $A_{\ell x}$, $A_{\ell y}$ and $A_{\ell z}$
%obtained with a linearly polarized photon beam and a polarized target 
%for the $\gamma n\to n' \gamma^*\hookrightarrow e^+e^-$ reaction.

\begin{center}
\begin{figure}[htbp]
\hspace*{-0.8cm}
\includegraphics[width=6.5cm,height=6cm]
{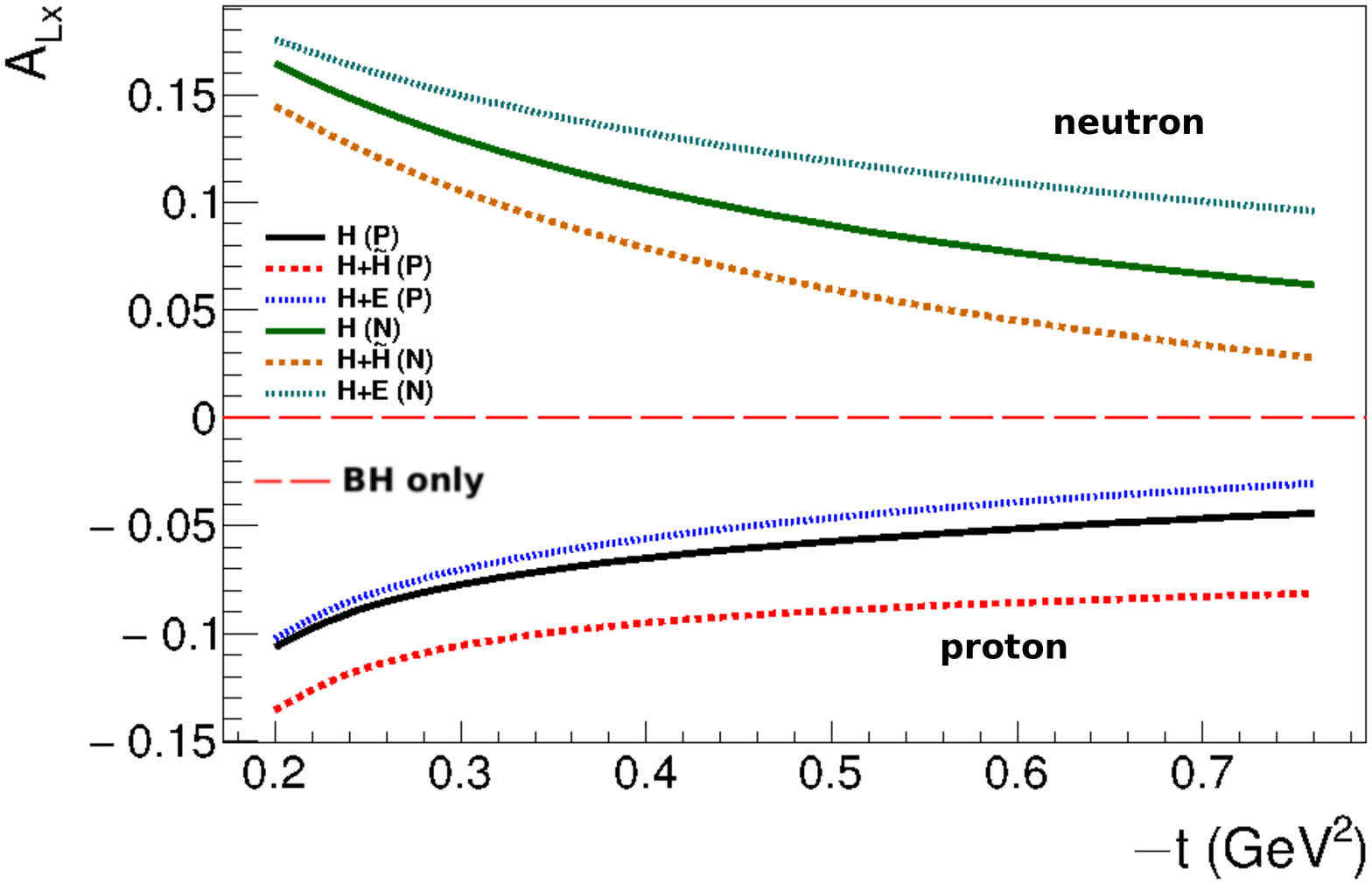}
\hspace*{-0.8cm}
\includegraphics[width=6.5cm,height=6cm]
{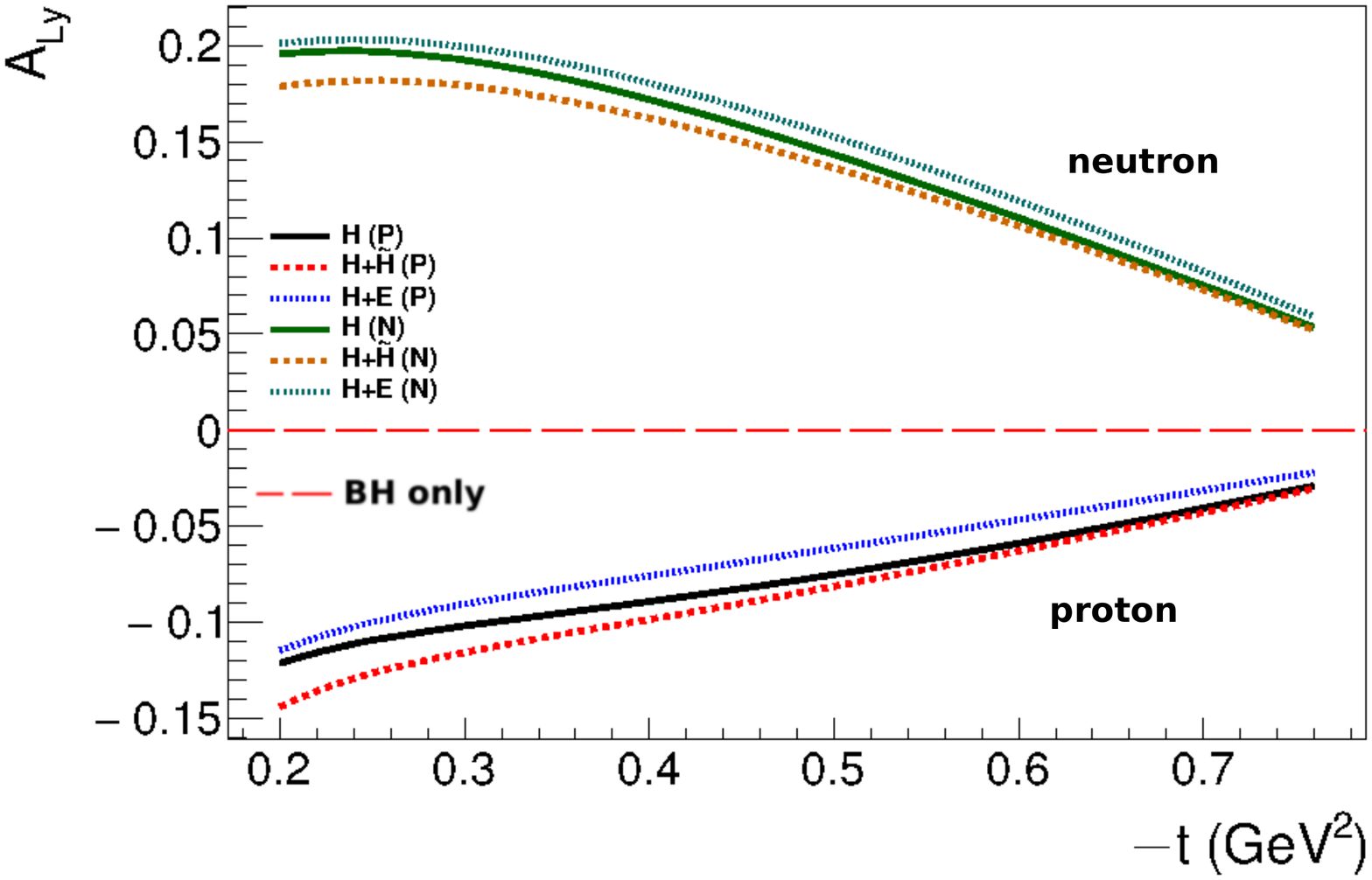}
\hspace*{-0.8cm}
\includegraphics[width=6.5cm,height=6cm]
{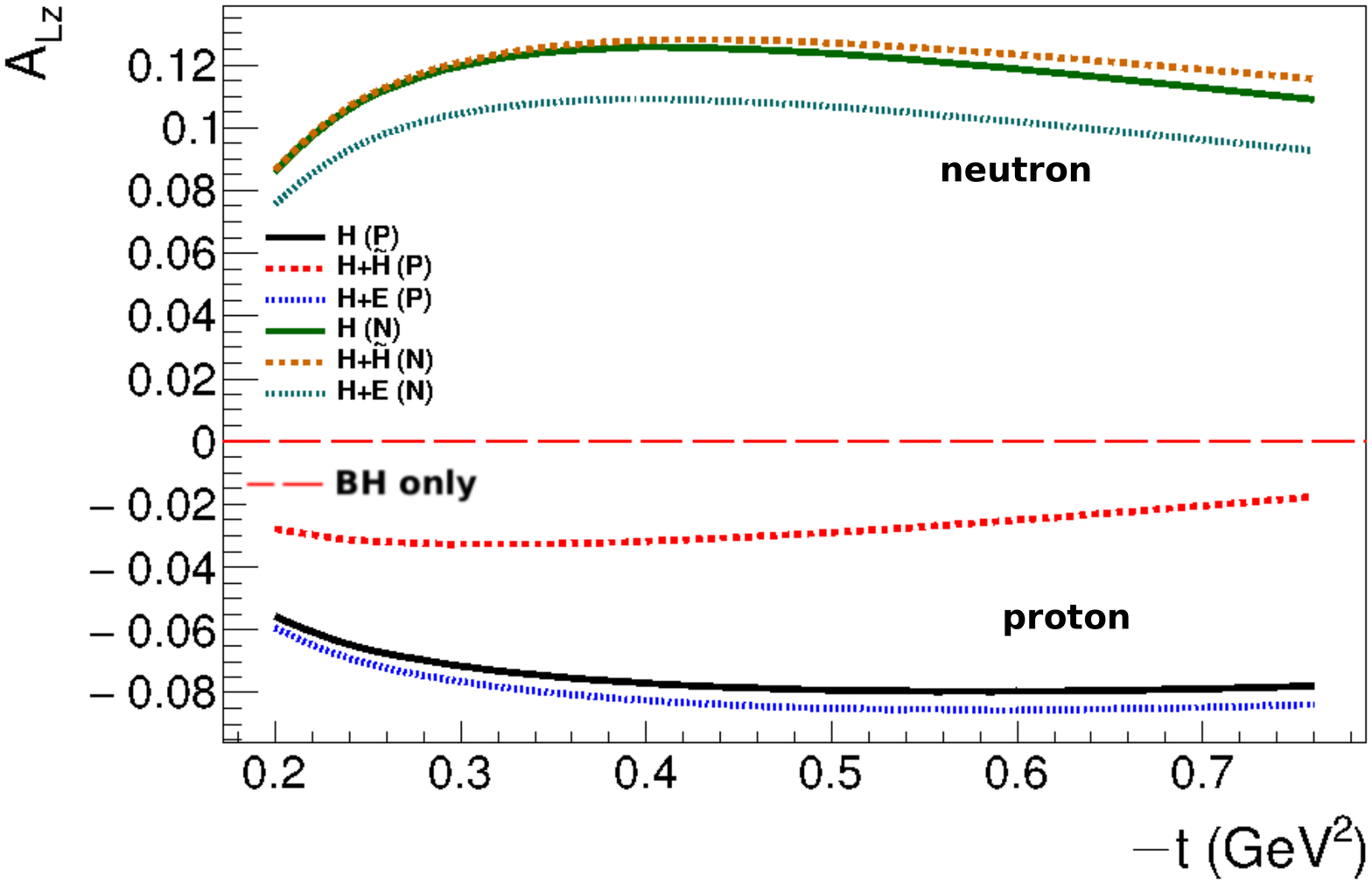}
\hspace*{-0.8cm}
\caption{The double-spin asymmetries $A_{\ell x}$ (left panel),  $A_{\ell y}$ (central panel) 
and $A_{\ell z}$ (right panel), respectively for
$\phi=90^\circ$, $\phi=0^\circ$ and $\phi=90^\circ$,
as a function of $-t$ for $\xi=0.2$, $Q'^2=7$ GeV$^2$ and for $\theta$ 
integrated over $[\pi/4,3\pi/4]$.
The results are shown for the neutron and proton target cases.}
\label{fig:BTSAlint}
\end{figure}
\end{center}

In Fig.~\ref{fig:BTSAlint}, we show the $t$-dependence of 
the three double spin asymmetries $A_{\ell x}$, $A_{\ell y}$ and $A_{\ell z}$
obtained with a linearly polarized photon beam and a polarized target 
for the neutron and proton target cases.
Like for the proton target case~\cite{Boer:2015hma},
the asymmetries have respectively a $\sin2\phi$, $\cos2\phi$ and $\sin2\phi$-like shape.
We therefore plot in Fig.~\ref{fig:BTSAlint} $A_{\ell x}$, $A_{\ell y}$ and $A_{\ell z}$ 
at $\phi=90^\circ$, $\phi=0^\circ$ and $\phi=90^\circ$ respectively, for 
our typical kinematics $\xi=0.2$, $Q'^2=7$ GeV$^2$,
$-t=0.4$ GeV$^2$ and $\theta$ integrated over $[\pi/4,3\pi/4]$.
This double-spin asymmetry is sensitive to the
imaginary part of the amplitude and therefore the BH alone does not
produce any signal. As expected,
the sign is opposite for the neutron and proton target cases.
We note, for the neutron target case, the particular sensitivity of $A_{\ell x}$ to the GPDs
$H^n$, $\tilde H^n$ and $E^n$ and of $A_{\ell z}$ to the GPDs $H^n$ and $E^n$.

\onecolumngrid

%\begin{center}
%\begin{figure}[htbp]
%\hspace*{-0.8cm}
%\includegraphics[width=6.5cm,height=6cm]
%{Figures2/BTSA_lin_px_vsphi_PN_0.pdf}
%\hspace*{-0.8cm}
%\includegraphics[width=6.5cm,height=6cm]
%{Figures2/BTSA_lin_py_vsphi_PN_1.pdf}
%\hspace*{-0.8cm}
%\includegraphics[width=6.5cm,height=6cm]
%{Figures2/BTSA_lin_pz_vsphi_PN_0.pdf}
%\hspace*{-0.8cm}
%\caption{The double-spin asymmetries $A_{\ell x}$ (left panel),  $A_{\ell y}$ (central panel) 
%and $A_{\ell z}$ (right panel) as a function of $\phi$ for $\xi=0.2$, $Q'^2=7$ GeV$^2$,
%$-t=0.4$ GeV$^2$ and for $\theta$ 
%integrated over $[\frac{\pi}{4},\:\frac{3\pi}{4}]$.}
%\label{fig:BTSAlinphi}
%\end{figure}
%\end{center}

\twocolumngrid

\begin{center}
\begin{figure}[htbp]
\hspace*{-0.8cm}
\includegraphics[width=6.5cm,height=6cm]{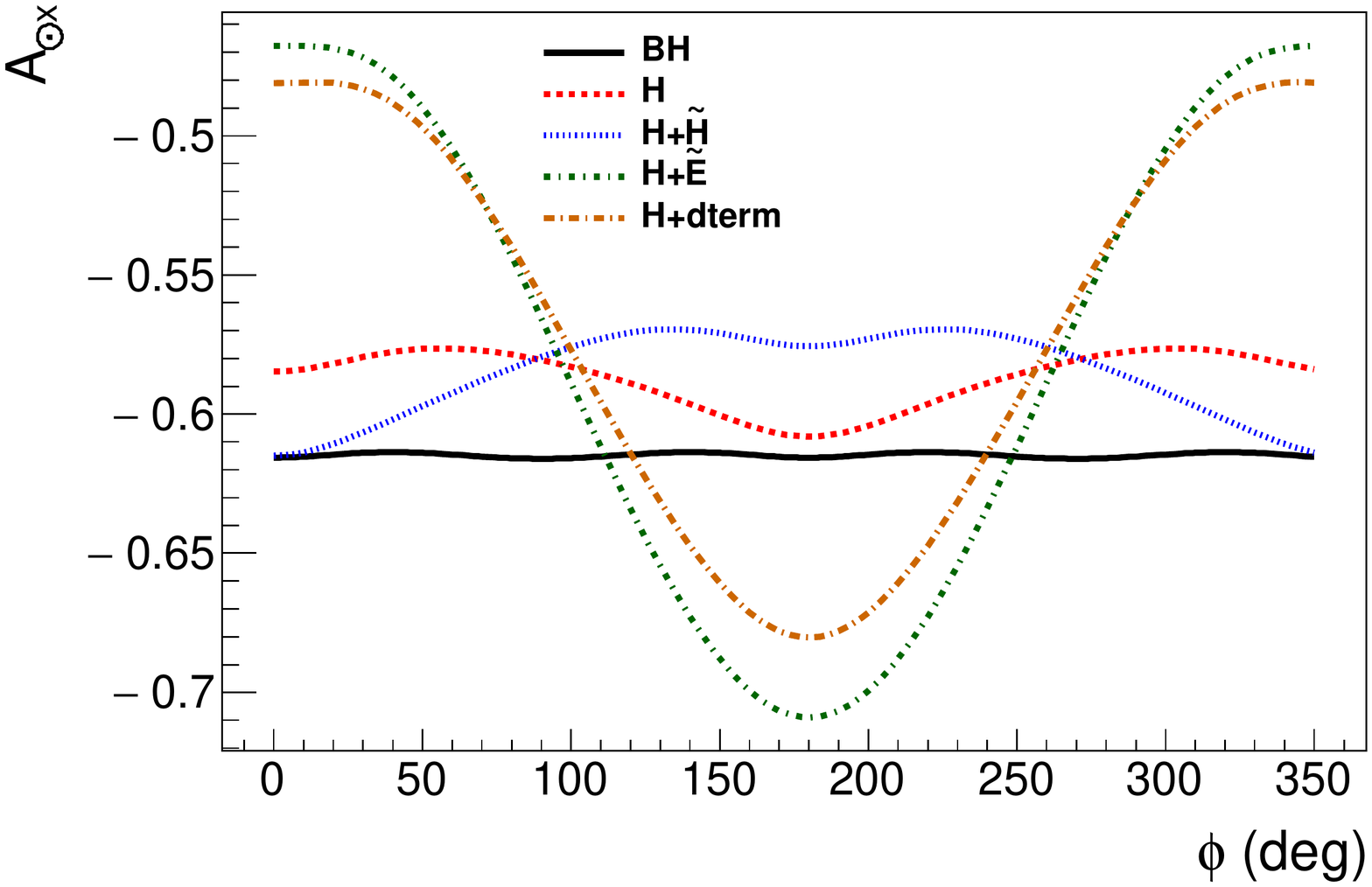}
\hspace*{-0.8cm}
\includegraphics[width=6.5cm,height=6cm]{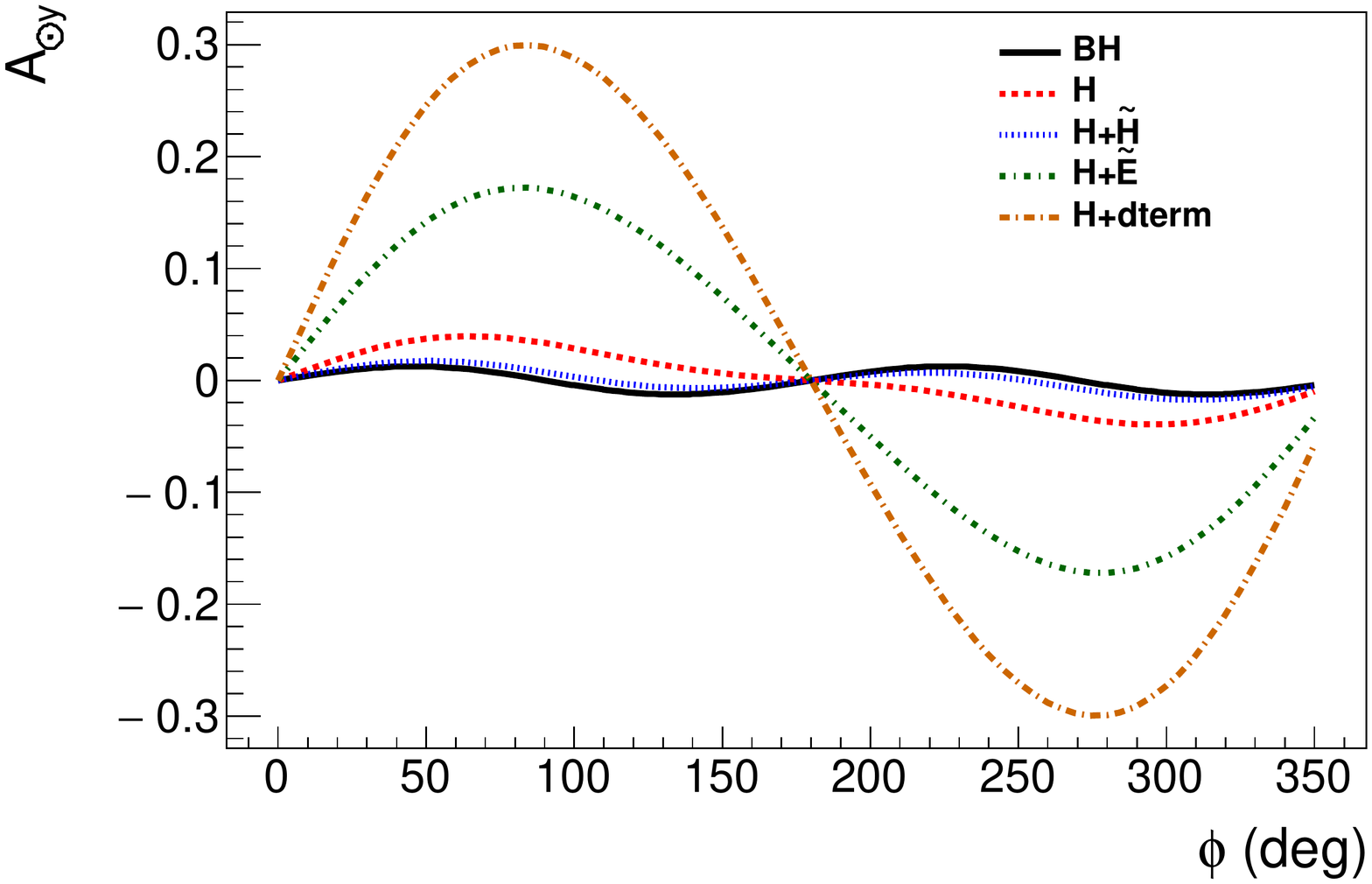}
\hspace*{-0.8cm}
\includegraphics[width=6.5cm,height=6cm]{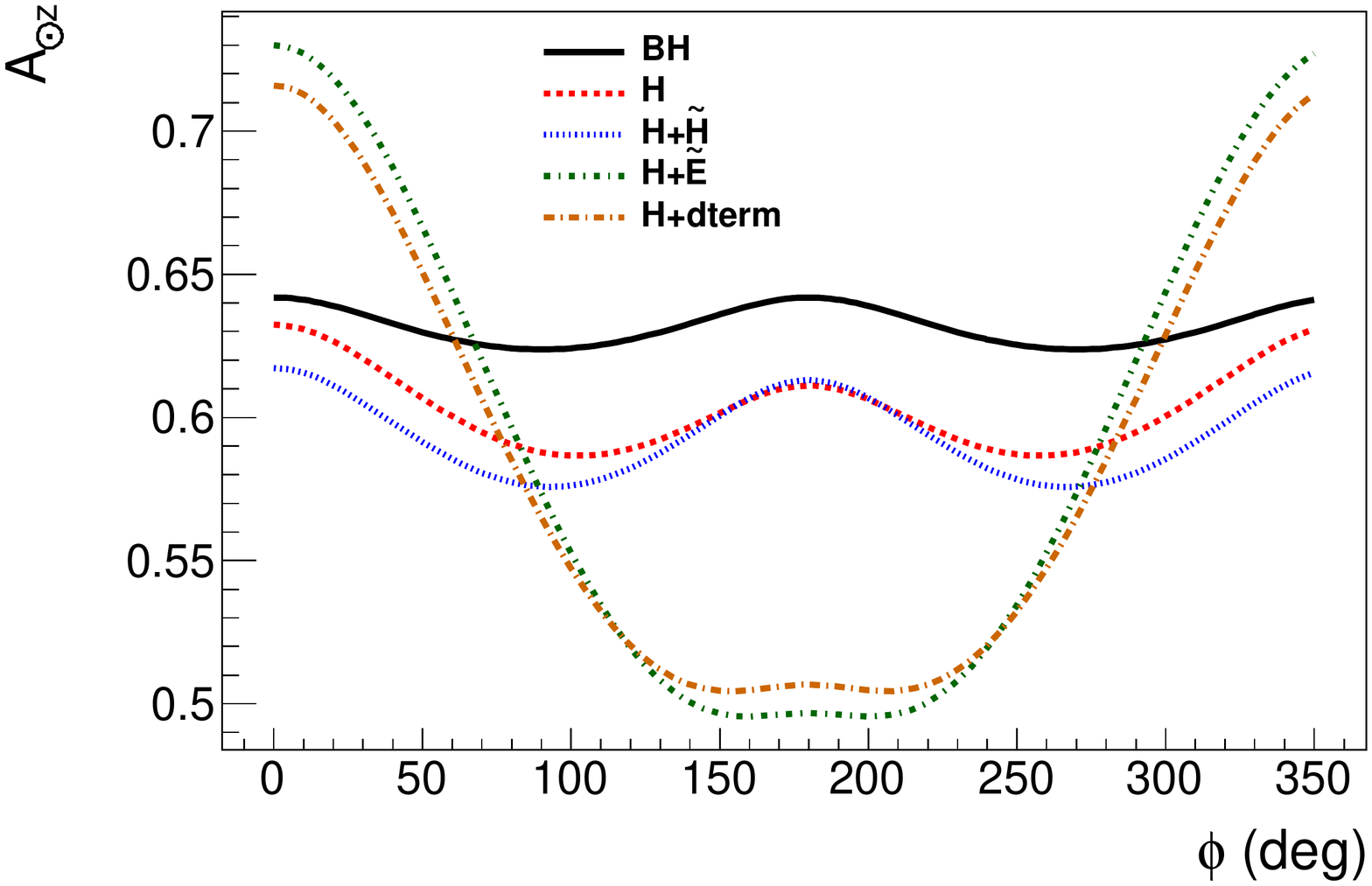}
\hspace*{-0.8cm}
\caption{The double-spin asymmetries $A_{\odot x}$ (left panel), $A_{\odot y}$ (central panel) 
and $A_{\odot z}$ (right panel) as a function of $\phi$ for $\xi=0.2$, $Q'^2=7$ GeV$^2$,
$-t=0.4$ GeV$^2$ and for $\theta$ 
integrated over $[\pi/4,3\pi/4]$.}
\label{fig:BTSAcircphi}
\end{figure}
\end{center}

In Fig.~\ref{fig:BTSAcircphi}, we show the $\phi$-dependence of 
the three double-spin asymmetries $A_{\odot x}$, $A_{\odot y}$ and $A_{\odot z}$
obtained with a circularly polarized photon beam and a polarized target 
for the $\gamma n\to n' e^+e^-$ process.
The kinematics is as before: $\xi=0.2$, $Q'^2=7$ GeV$^2$,
$-t=0.4$ GeV$^2$ and $\theta$ has been
integrated over $[\pi/4,3\pi/4]$.
We show here the $\phi$-dependencies as they can be quite 
intricate and disparate.
This double spin asymmetry is sensitive to the
real part of the amplitude and therefore the BH alone can
produce a strong signal by itself. The TCS can
change significantly and in various ways the asymmetries w.r.t. BH alone,
making this observable very sensitive to various GPD contributions.
In particular, one notes the sensitivity to $\tilde E^n$.
%We don't notice any strong sensitity to $E^n$.
The most important TCS influence appears in $A_{\odot y}$
where the asymmetry amplitude can change from only a few percent
for BH alone to about 30\% for a particular GPD configuration. 
%One notices in particular the important sensitivity of these three
%double-spin asymmetries to the ansatz taken for the
%$t$-dependence of $H^n$ (Reggeized vs factorized). We also noticed
%this feature for the proton target case~\cite{Boer:2015hma}.

\begin{center}
\begin{figure}[htbp]
\hspace*{-0.8cm}
\includegraphics[width=6.5cm,height=6cm]
{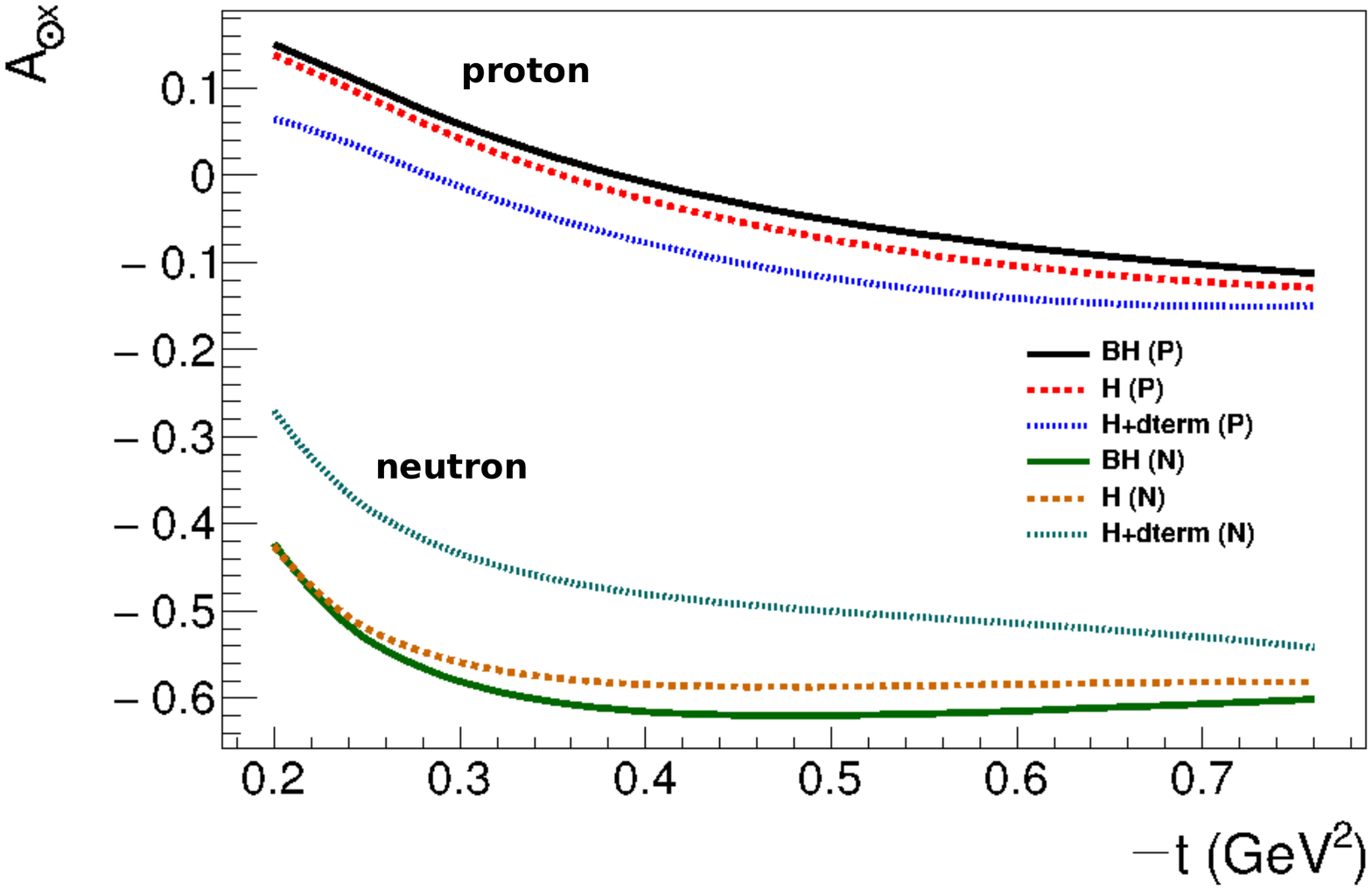}
\hspace*{-0.8cm}
\includegraphics[width=6.5cm,height=6cm]
{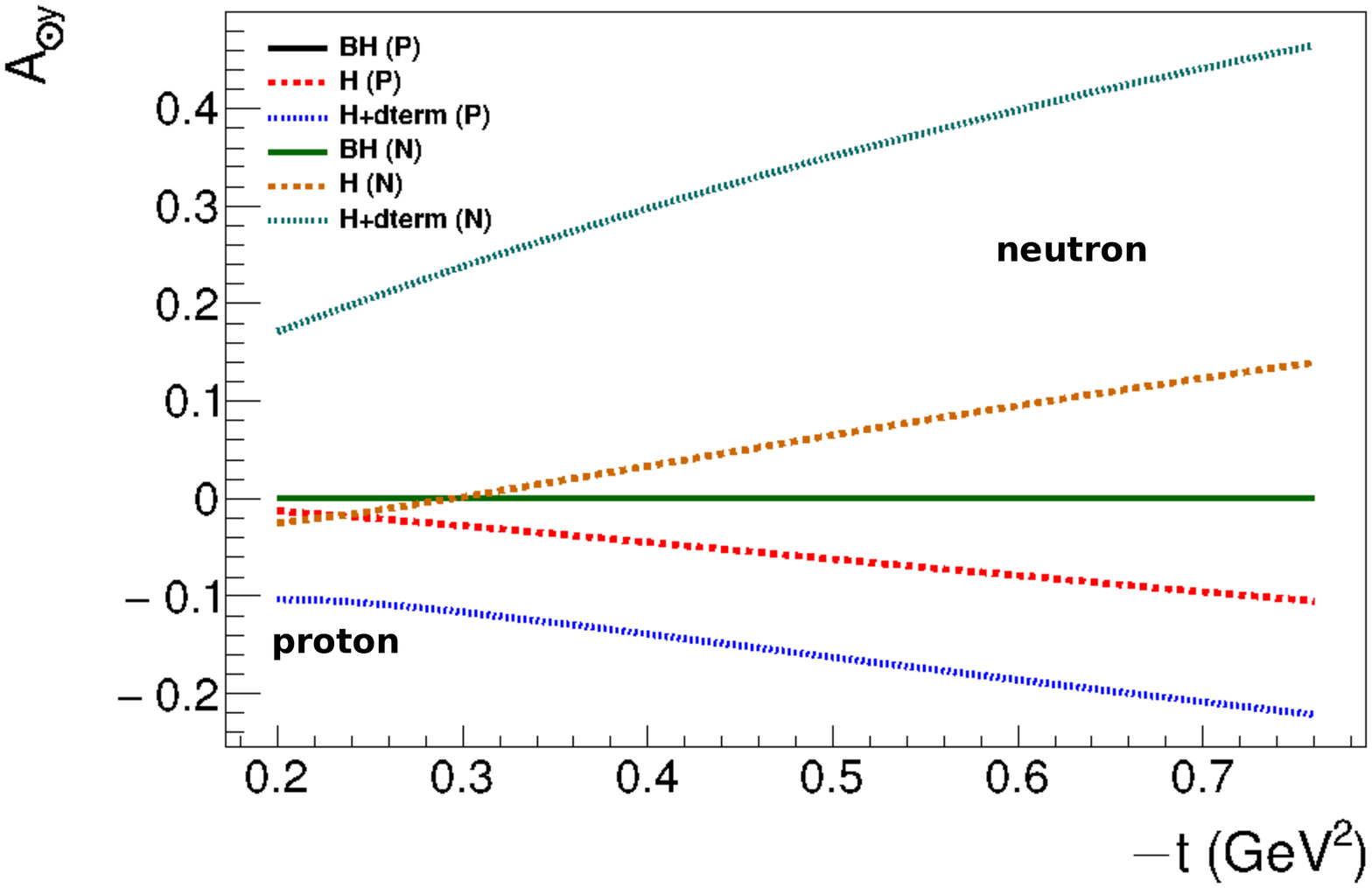}
\hspace*{-0.8cm}
\includegraphics[width=6.5cm,height=6cm]
{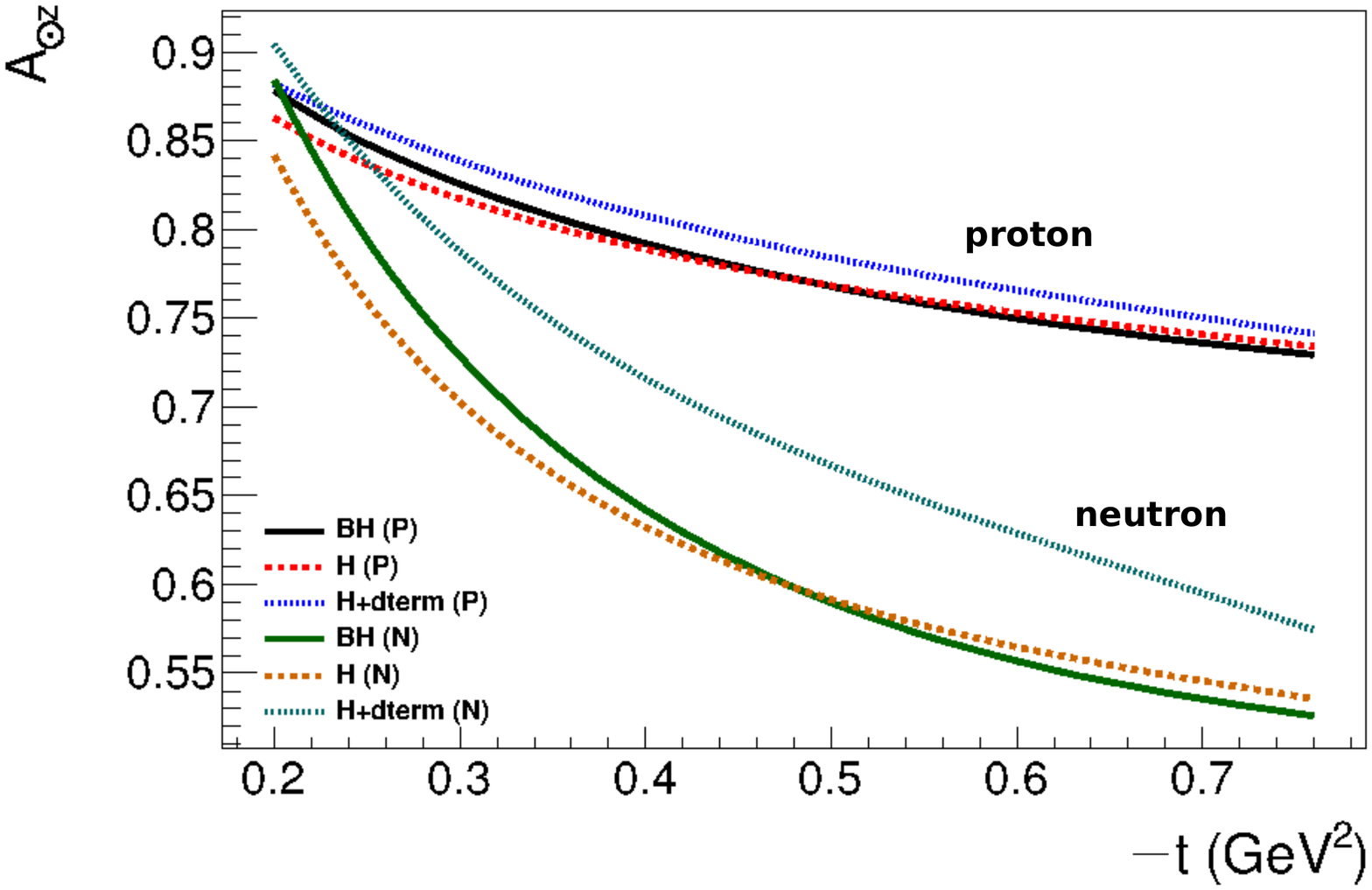}
\hspace*{-0.8cm}
\caption{The $A_{\odot x}$ (left panel),  $A_{\odot y}$ (central panel) 
and $A_{\odot z}$ (right panel)
asymmetries, respectively for
$\phi=0^\circ$, $\phi=90^\circ$ and $\phi=0^\circ$, as a function of $t$ 
for $\xi=0.2$, $Q'^2=7$ GeV$^2$, $-t=0.4$ GeV$^2$ and for $\theta$ 
integrated over $[\pi/4,3\pi/4]$.
Calculations are done for the neutron and proton target cases.}
\label{fig:BTSAcirct}
\end{figure}
\end{center}

We finally show in Fig.~\ref{fig:BTSAcirct} the
$t$-dependence of $A_{\odot x}$, $A_{\odot y}$ and $A_{\odot z}$ at 
$\phi=90^\circ$, $\phi=0^\circ$ and $\phi=90^\circ$ respectively,
for the neutron and proton target cases. 
%We note, for the neutron target case, the particular sensitivity of $A_{\ell x}$ to the 
%GPDs $H^n$ and $E^n$ and of $A_{\ell z}$ to the GPD $E^n$.

\onecolumngrid

\twocolumngrid

\section{Conclusions}

In summary, we have presented a first phenomenological study of 
the $\gamma n\to n' e^+e^-$ reaction in the handbag approach. Using
the GPDs from the VGG model, we calculated the unpolarized cross
section and all the beam and target single and double spin 
asymmetries of the process for typical kinematics of JLab at 12 GeV.
We showed the sensitivities of these observables to various GPDs.
In particular, we highlighted the sensitivity of the circularly polarized
beam spin asymmetry to the elusive GPD $E$ (of the neutron), which is
of special interest for the study of the nucleon spin decomposition.

The cross section of the $\gamma n\to n' e^+e^-$ process is 
only a factor 2 below the $\gamma p\to p' e^+e^-$, for which experimental
proposals have been approved at JLab, and some asymmetries are sizeable,
even more than for the proton in some cases. Thus,
the study of TCS on the neutron appears feasible experimentally and
promises to bring new important constraints on GPD physics, in particular
on the GPD $E^n$ and, more generally, the flavor separation of GPDs.

\section*{Acknowledgments}

The work of M.V. is supported by the Deutsche Forschungsgemeinschaft DFG 
through the Collaborative Research Center ``The Low-Energy Frontier of the
Standard Model" (SFB 1044) and the Cluster of Excellence ``Precision Physics, Fundamental
Interactions and Structure of Matter" (PRISMA). 
M. G. and M. V. are also supported by the Joint Research Activity ``GPDex" of the
European program Hadron Physics 3 under the Seventh Framework Programme of the European Community.
M.B. and M.G. also benefitted from the GDR 3034 ``PH-QCD" and the ANR-12-MONU-0
008-01 ``PARTONS" support.

%########################################################################
% Bibliographie
%########################################################################

%%%%%%%%%%%%%%%%%%%%%%%%%%%%%%%%%%%%%%%%%%%%%%%%%%%%%%%%%%%%%%%%%%%%%%%%%%%%%%%%%%%
%%%%%%%%%%%%%%%%%%%%%%%%%%%%%%%%%%%%%%%%%%%%%%%%%%%%%%%%%%%%%%%%%%%%%%%%%%%%%%%%%%%

\end{document}